\documentclass[10pt,journal,compsoc]{IEEEtran}
\usepackage{amssymb}
\usepackage{amsthm}
\usepackage{amsmath}
\usepackage{booktabs} 
\usepackage{color}
\usepackage{xcolor}
\usepackage{tikz}
\usetikzlibrary{fit,positioning}
\usepackage{graphics}
\usetikzlibrary{arrows}
\usepackage{algorithmic}
\usepackage{algorithm}
\usepackage{subcaption}
\usepackage[switch]{lineno}
\usepackage{float} 
\usepackage{hyperref}
\usepackage{tikz}
\usetikzlibrary{arrows,positioning,shadows}
\usepackage{standalone}
\makeatletter
\def\mathcolor#1#{\@mathcolor{#1}}
\def\@mathcolor#1#2#3{%
	\protect\leavevmode
	\begingroup
	\color#1{#2}#3%
	\endgroup
}
\makeatother
\ifCLASSOPTIONcompsoc
  \usepackage[nocompress]{cite}
\else
  \usepackage{cite}
\fi
\ifCLASSINFOpdf

\else
 \fi

\begin{document}

\title{Detection of Community Structures in Networks with Nodal Features based on Generative Probabilistic Approach}
%

\author{Hadi~Zare,
        Mahdi~Hajiabadi, Mahdi Jalili,~\IEEEmembership{Senior Member,~IEEE}
\IEEEcompsocitemizethanks{\IEEEcompsocthanksitem Hadi Zare and Mahdi Hajiabadi are with the Faculty of New Sciences and Technologies, University of Tehran, Tehran, Iran. Mahdi Jalili is with the School of Engineering, RMIT University, Melbourne, Australia. Mahdi Jalili is supported by Australian Research Council through project No. DP170102303.\protect\\
E-mail: h.zare@ut.ac.ir, m.hajiabadi@ut.ac.ir, mahdi.jalili@rmit.edu.au
}
\thanks{Manuscript received ; revised .}}

\markboth{Journal of \LaTeX\ Class Files}%
{Shell \MakeLowercase{\textit{et al.}}: Bare Demo of IEEEtran.cls for Computer Society Journals}

\IEEEtitleabstractindextext{%
\begin{abstract}
Community detection is considered as a fundamental task in analyzing social networks. Even though many techniques have been proposed for community detection, most of them are based exclusively on the connectivity structures.  However, there  are node features in real networks, such as gender types in social networks, feeding behavior in ecological networks, and location on e-trading networks, that can be further leveraged with the network structure to attain more accurate community detection methods. We propose a novel probabilistic graphical model to detect communities by taking into account both network structure and nodes' features. The proposed approach learns the relevant features of communities through a generative probabilistic model without any prior assumption on the communities. Furthermore, the model is capable of determining the strength of node features and structural elements of the networks on shaping the communities. The effectiveness of the proposed approach over the state-of-the-art algorithms is revealed on synthetic and benchmark networks.

\end{abstract}

}

\maketitle

\IEEEdisplaynontitleabstractindextext

\IEEEpeerreviewmaketitle

\IEEEraisesectionheading{\section{Introduction}\label{sec:introduction}}                                                             
\IEEEPARstart{A} network is a set of inter-connected items  with a powerful mathematical basis for modeling many real-world systems, such as the  Internet, World Wide Web, and transportation networks \cite{newman2003structure}. Detection of communities as the main hidden structures of networks has attracted the interests of researchers from the early stages of the the appearance of network science \cite{fortunato2016community}.  A community in a network is defined as a set of nodes with intense intra community connections  while having sparse inter community links.  Nodes within the same community are likely to share common properties and play similar actions \cite{coscia_classification_2011}. The role of community structures in the functional modules of the networks has been applied on a wide range of fields including spammers identification in online social networks \cite{bhat2013}, image clustering \cite{okuda2017}, and detection the neural units dense modules \cite{garcia_applications_2018}.\par 
Much effort has been carried out on various aspects and assumptions about communities  while focusing primarily on the connectivity information. Algorithms to discover non-overlapping communities are generally aimed at partitioning the network into sub-networks which are densely connected internally, while weakly connected externally. Examples of such algorithms are  graph partitioning \cite{girvan2002community},  hierarchical agglomeration algorithm \cite{clauset2004finding}, optimization based methods \cite{ blondel2008fast} and many variants of spectral clustering method \cite{qin2013regularized}. On the other hand, several methods have been proposed to discover overlapping communities, such as mixed membership stochastic block-models \cite{airoldi2008mixed}, map equation framework based on probabilistic flows (\emph{InfoMap}) \cite{rosvall_maps_2008}, label propagation (\emph{Fast-Greedy}) \cite{gregory_finding_2010}, nonnegative matrix factorization (\emph{BigClam}) \cite{yang2013overlapping}, modularity based optimization (\emph{COMBO}) \cite{sobolevsky_general_2014},  tracking the evolution of online social networks \cite{bhat2015}, neighborhood seed expansion \cite{whang2016}, a unified approach on detection of general community structures \cite{hajiabadi_iedc2017}, and asymmetric triangle cuts \cite{rezvani2018}.   
\par
Indeed, the community detection can be considered as an ill-posed hard unsupervised learning problem. There are node (or edge) features that can be effectively used to provide better community structures \cite{peelGroundTruthMetadata2017a, chakraborty2018}. On the one hand, it is known that significant correlation exists between community structure and node features, hereafter called as \emph{``features''}, in a variety of real networks \cite{fortunato2016community}. On the other hand, most of the well-known approaches are based on applying only one of these two information sources.  Exploiting features in the community detection process could  yield better results. Moreover, it is shown that there is strong dependency between the communities and features in some real networks  \cite{newman2016structure}.
Recently, some researchers  have addressed the community detection  using network structure coupled with the features, such as  single-assignment clustering heuristic  \cite{ruan2013efficient}, topic derived models \cite{sun2012relation}, generative model (\emph{CESNA}) \cite{yang2013community}, Bayesian Graph Clustering (\emph{BAGC}) \cite{xu2012model}, and Expectation-Maximization (EM) approach \cite{newman2016structure}. However, most of these approaches are somewhat sensitive to correctness of the model specification. \par
In this paper, we propose a novel graphical model to find communities through a  probabilistic approach. The proposed model provides the level of correlations between communities and features that could be used to select the suitable divisions of the network as well as the appropriate features. The summary of our contributions in this work are as follows,
\begin{itemize}
	\item We propose a graphical model to form the relation between the communities and features
	\item We investigate the correlation between the community structures and features based on the learned model
	\item We extract communities for a general class of networks through exploiting features
	\item We introduce a novel approach on the influence of  features on the community structures
\end{itemize}
\par
The paper is organized as follows.  Section \ref{sec2} presents the related works and motivation of the proposed approach. Section \ref{sec3} introduces the elements of the proposed model. Section \ref{sec4} describes the statistical learning of  the model parameters. Section \ref{sec5} represents the experimental results on benchmark real network dataset. Section \ref{sec6} presents a case study to illustrate the proposed approach. Section \ref{sec7} concludes the paper with suggestions for further works on this field.  
\section{Related works and motivation}
\label{sec2}
\subsection{Related works}
The role of nodal features on different aspects of  network modeling are considered in a variety of works such as the missing nodes prediction via non-parametric Bayesian inference \cite{hric2016}, finding k-truss subgraphs with the aid of features \cite{huang2017}, and network approach on topic modeling \cite{gerlach2018}. \par 
On community detection with nodal features, there are generally two types of techniques, model-free methods and generative models. Like the structure based algorithms with some optimality criteria to detect communities such as the modularity based methods \cite{newman2006modularity, blondel2008fast, chen2014} and label propagation \cite{lu2019}, such model-free methods are proposed to exploit the features including structure mining \cite{silva2012mining}, simulated annealing \cite{cheng2011clustering}, Joint Community Detection Criterion (\emph{JCDC}) \cite{zhang_community_2016},  Semidefinite Programming (\emph{SDP}) \cite{yan2016convex}, and Covariance Assisted Spectral Clustering (\emph{CASC}) \cite{binkiewicz_covariate-assisted_2017}. Most methods in this category exploit features in the same way without considering the relationship between them and communities. \par 
Generative models were initially introduced on connectivity based community detection including affiliation graph model \cite{yang2012a}, matrix factorization \emph{BigClam} \cite{yang2013overlapping}, Bayesian community detection \cite{pas2018}, and nonparametric probabilistic model by conducting random walks \cite{zhu2019}. Feature based generative models on community extraction have been proposed in some works such as topic modeling \cite{sun2012relation}, \emph{CESNA} \cite{yang2013community}, and stochastic block model \cite{newman2016structure}. 
In \cite{yang2013community}, a generative model  was introduced to consider just the influence of community structures on features.
The modified stochastic block model aligned with the features is modified in \cite{newman2016structure} to reveal the efficacy of each feature on community structures by employing the Expectation-Maximization inference stage. 
On the one hand, most of the model-free feature based methods suffer from the dependency to multiple tuning parameters such as \emph{JCDC} \cite{zhang_community_2016}, and \emph{CASC} \cite{binkiewicz_covariate-assisted_2017}. On the other hand, the generative feature based models on extraction of communities have some problems including the model sensitivity on the presumed graphical representation of the features and communities \emph{CESNA} \cite{yang2013community}, and modeling a correlation of single feature with the community structure at a time\cite{newman2016structure}. 
\subsection{Motivation}
In general, there are two paradigms in constructing the effect of features on the community structure: \textbf{(i)} the assortative features like age, sexuality, race and overall personal user's attributes having significant influence on the formation of communities, and \textbf{(ii)}  the community generative features like education, living place, office location and user's interests in social networks which is imitated from the community structure.   
\begin{figure}[t!]
	\centering
	\begin{minipage}{0.25\textwidth}
		\centering
		\includegraphics[width=0.5\linewidth]{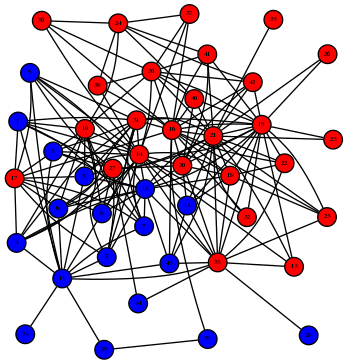}
		\subcaption{}
		\label{SubFig:Weddell}
	\end{minipage}%
	\begin{minipage}{0.25\textwidth}
		\centering
		\includegraphics[width=0.5\linewidth]{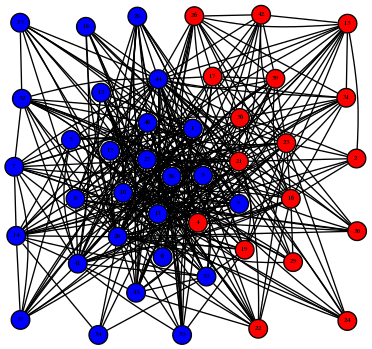}
		\subcaption{}
		\label{SubFig:WorldTrade}
	\end{minipage}
	\caption{The influence of features on communities in two real networks, \textbf{(a)}: \emph{Weddell Sea}  with \emph{feature} ``Environment'' with two categories, "Pelagic" (Blue) and "Benthic" (Red). \textbf{(b)}: \emph{World Trade} with \emph{feature}  "continent" where each node(country) belongs to "Asia" (Blue) or "Europe" (Red). }
	\label{fig:FeatImpaceComm}
\end{figure}
To clarify the effect of features on the formation of community structure, we consider two real-world networks. Figure \ref{fig:FeatImpaceComm} (a) shows a snapshot of the \emph{Predator-Prey} network where each node represents a unique marine creature of the Weddell Sea and the color of each node shows its living environment , ``Pelagic'' or ``Benthic'' \cite{jacob2011role}. Figure \ref{fig:FeatImpaceComm} (b) shows a snapshot of  \emph{World Trade} network,  where each node (country) belongs to "Asia" or "Europe" \cite{de2011exploratory}. As it can be seen, the features provide a useful insight on discovering the more likely community structures on these case studies. 

Our primary aim in this work is to study how  to extract community based on two different type of features, assortative and generative,  which has not been considered in the earlier works. We propose a principled graphical model via the “division of features into assortative and generative features” to construct a general approach to deal with the community detection problem.  It is assumed that each community can be formed by causal relationship of assortative features and the \emph{community} generative features which are influenced within the community structure.
The community generative features is called as \emph{generative features} for simplicity.
\begin{figure}[t]
	\centering
	\includegraphics[width=0.5\linewidth]{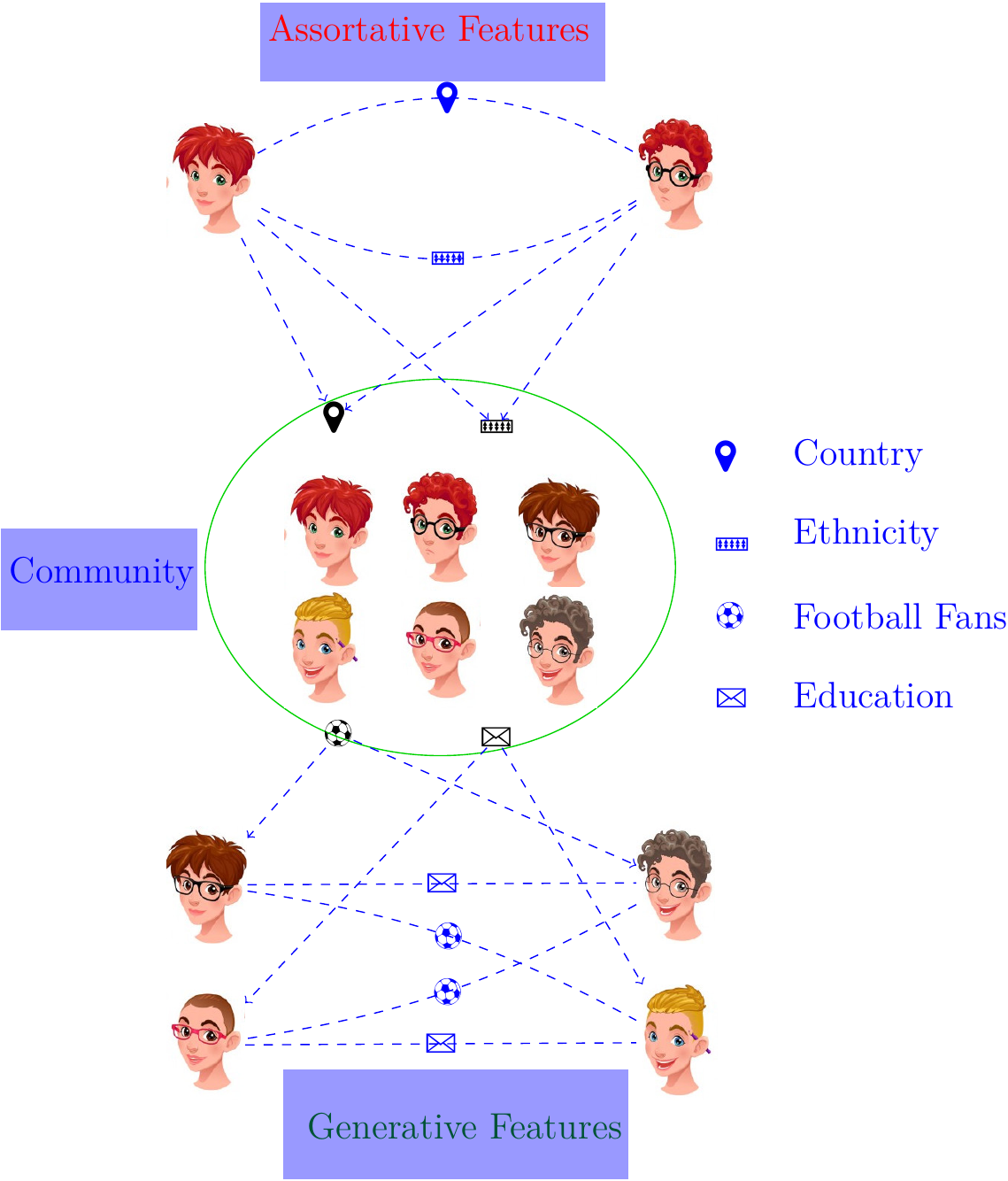}
	\caption{The connection between features and community structure where \emph{gender} and \emph{ethnicity} are \emph{assortative features} with causal impact on the community strucutre,  and \emph{football} and \emph{education} are generative features which are generated by the community structure.}
	\label{fig:SmallScheme}
\end{figure}

The causal relationship is represented in a graphical model to encode the main elements of the community formation  through two main sources:  connectivity structure and features. The schematic representation of  features and their relations on the community structures are depicted in Figure \ref{fig:SmallScheme}.  The proposed approach is designed based on the different types of features in the probabilistic generative model. Moreover, it can also model the joint features altogether  unlike most of ad-hoc based earlier methods \cite{newman2016structure, zhang_community_2016}. 
The parameters of the proposed model are learned through a likelihood based approach. Furthermore, the dependency of features on the community structure is computed through the learning phase that can be applied to infer the main ingredients on the construction of network communities. 


\begin{figure}[t!]
	\centering
	\includegraphics[width=1\linewidth]{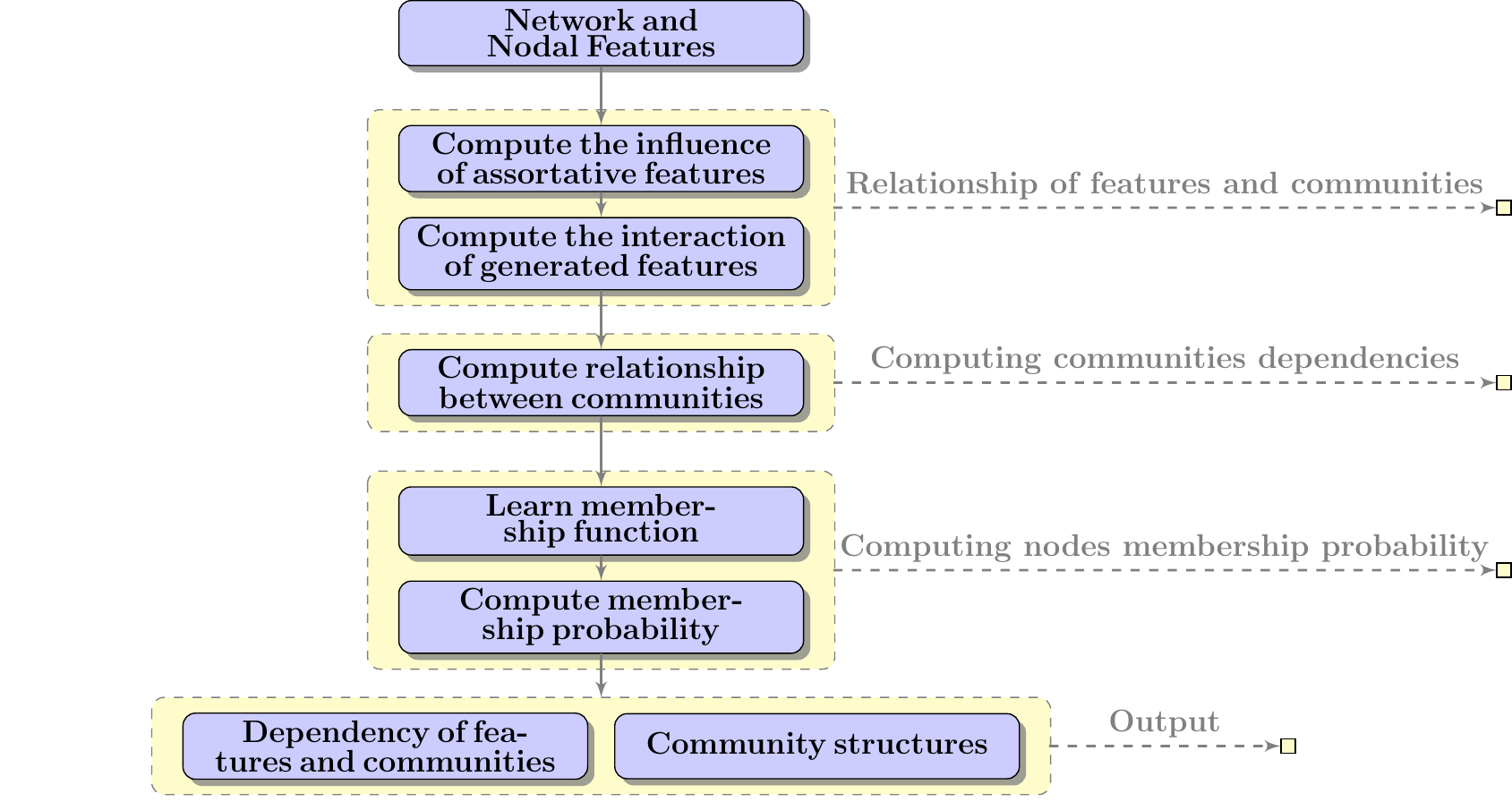}
	\caption{\label{fig:SchemePaper}  {The flow-graph of the proposed method}}
\end{figure}

\section{Elements of the proposed approach}
\label{sec3}
Let's assume a network representation as $G = (V, E)$ where $V$ and $E$  denote nodes and edges in graph $G$. Furthermore, we have a set of features on nodes as $Features = S\cup F$, which comprises two disjoint subsets  $S$  for ``assortative features'' and $F$ for ``generative features'' based on the dataset contextual information. The assortative features represent the personal attributes like race and age, and the generative features represent features like education, sports and users' interests, see Figure \ref{fig:SmallScheme} for further details. \par
The proposed approach consists of three main steps, which is depicted in Figure \ref{fig:SchemePaper}. Initially, the features are divided into assortative and generative.  {The first main step is to measure the dependency of features \textcolor{black}{towards each community}. In this step, the influence of assortative  and generative features on the community formation is computed. The first main step is to measure the dependency of features towards each community. In this step, the influence of assortative  and generative features on the community formation is computed. The relationship between communities is the second component of the proposed algorithm. The next important step is computing  the probability membership function for each node, which is performed by statistical learning of the main parameters and iteratively updating the initial membership function to result in the final ones. Output of the algorithm is twofold: the dependency weights of features to the community structures that is informative for interpretation of the attained results, and the community structure of the network.

	\subsection{Description of graphical model}
	Here, we describe the details of proposed community detection model. The proposed model  provides a generative framework among the main factors in a graph structure to detect the community structures in a network. 
	The graphical model is constructed based on main factors namely community membership $M$, community interaction matrix $\beta$,  assortative features $S$, generative features $F$,  the influence of assortative features on communities denoted by  $I$  and the interaction of communities with features denoted by $W$. 
	\begin{figure}[t!]
		\centering
		\begin{minipage}{0.50 \textwidth}
			\centering
			\includegraphics[width=.4\linewidth]{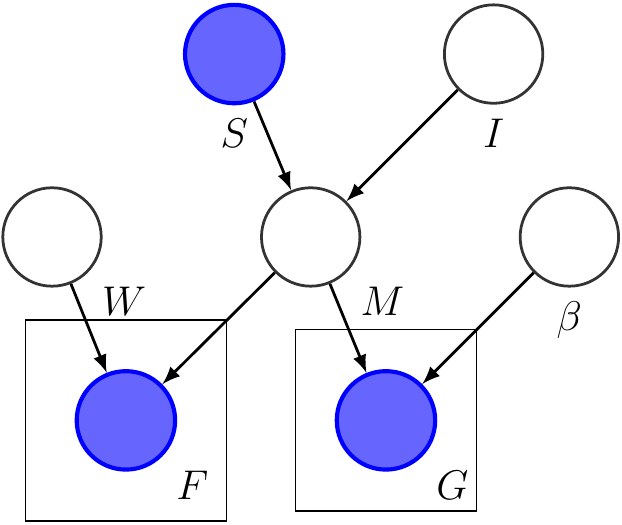}
		\end{minipage}%
		\\
		\begin{minipage}{0.50\textwidth}
			\begin{table}[H]
				\centering
				\small
				\resizebox{1\textwidth}{!}{
					\begin{tabular}{lcc}
						\toprule[1.5pt]
						
						Node &  Type & Description\\ 
						\midrule 
						$M$ & Hidden Variable & Membership function of each node\\
						$I$ & Weight Factor & Correlation level between communities and assortative features\\
						$W$ &Weight Factor & Correlation level between communities and generative features \\
						$\beta$ & Weight Factor & Level of interactions between communities \\
						$G$ & Observation & Input network \\
						$F$ & Observation & Generative features\\
						$S$ & Observation& Assortative features\\
						\bottomrule[1.5pt]    
					\end{tabular} 
				}
			\end{table}
			
		\end{minipage}
		\caption{The description of main elements of proposed methodology in a graphical model}
		\label{fig:ProposedMehodologyGraphicalModel}
	\end{figure}
	Figure \ref{fig:ProposedMehodologyGraphicalModel} represents the graph structure of the proposed model. \par
	The probability of creation of  an edge between a pair of nodes is directly related to their communities and the interaction levels between communities based on a probabilistic generative approach. Particularly, it is assumed that two nodes $u$ and $v$ are connected by considering the following probability, 
	\begin{equation}
	P ((u,v)\in E) =  1 - \exp(-M_u^T \beta M_v) 
	\label{eq:EdgeProbability}
	\end{equation}
	where $M_u$ and $M_v$ are non-negative membership functions of nodes $u$ and $v$ toward each community and $\beta$ represents the probability matrix of interactions between different communities. If nodes $u$ and $v$ share more communities or belong to communities $c_i$ and  $c_j$  with high level of interaction ($\beta_{c_i,c_j}$) among them, their tendency to establish an edge will be increased.
	Hence, nodes $u$ and $v$ do not share a connection with the following probability, 
	\begin{equation}
	P ((u,v)\notin E) = 1 - P((u,v) \in E)  =  \exp(-M_u^T \beta M_v) 
	\label{eq:EdgeNotProbability}
	\end{equation}
	Based on the generative probabilistic process between any pair of nodes $(u,v)$, each pair of nodes are independently distributed as Bernoulli distribution.  \textcolor{black}{Therefore, each element  $A_{uv}\in \{0,1\}$ of the adjacency matrix is generated according to the following generative approach,
		\begin{IEEEeqnarray}{rcl}
			P_{uv} &=&  1 - \exp(-M_u^T \beta M_v) \nonumber\\
			A_{uv} & \sim & Bernoulli(P_{uv})
	\end{IEEEeqnarray} }
	The graphical model in Figure \ref{fig:ProposedMehodologyGraphicalModel} indicates that the generative features $F$ is conditionally dependent on variables $M$ and $W$,  which is assumed to be parametrized through the following sigmoid probabilistic function,
	\begin{equation}
	P(F_{uk} = 1) = \frac{1}{1 + \exp(-\sum_{i=1}^{K}M_{uc_i}W_{kc_i})}
	\label{eq:FeatProbability}
	\end{equation}
	where $F_{uk} = 1$ denotes the property of node $u$ to have the $k$th feature, $M_uc_i$ and $W_kc_i$ denote the membership of node $u$ in community $c_i$ and the interaction between the $k$-th feature and community $c_i$. In summary, we assume that $F_{uk}$ follows the Bernoulli distribution in the following way: 
	
	\begin{equation}
	P(F_{uk} = 0) = \frac{\exp(-\sum_{i=1}^{K}M_{uc_i}W_{kc_i})}{{1 + \exp(-\sum_{i=1}^{K}M_{uc_i}W_{kc_i})}}
	\label{eq:FeatNotProbability}
	\end{equation}
	Furthermore, communities are  influenced by the assortative features $S$ and its weight parameters $I$ in the graph structure. In a similar way, the community membership of each node is estimated by,
	\begin{equation}
	M_{uc_i} = \frac{1}{1 + \exp(-\sum_{j\in I}I_{jc_i}S_{uj})}
	\label{eq:ProposedMembership}
	\end{equation}
	\section{Parameters learning}
	\label{sec4}
	Here, we consider the learning and inference stage on the proposed probabilistic model.
	The probability distribution on the observed variables $M$ and $F$ is written as,
	\begin{equation}
	\label{eq:jointdis}
	P(G, F| M,\beta, I, W, S) = P(G| M, \beta, I, S)P(F|M, W)
	\end{equation}
	The likelihood function is calculated based on the model configuration as follows,
	\begin{IEEEeqnarray}{rCl}
		L(\theta) &=& \prod_{(u,v)}\big(1 - \exp(-M_u^T \beta M_v)\big)^{a_{uv}}\big(\exp(-M_u^T \beta M_v)\big)^{1-a_{uv}}\IEEEnonumber\\
		&\times& \prod_{u}\prod_{k\in F} \Big(\frac{1}{1 + \exp(-\sum_{c_i}M_{uc_i}W_{kc_i})}\Big)^{F_{uk}}\IEEEnonumber\\
		&\times& \Big(\frac{\exp(-\sum_{c_i}M_{uc_i}W_{kc_i})}{1 + \exp(-\sum_{c_i}M_{uc_i}W_{kc_i})}\Big)^{1-F_{uk}}
		\label{eq:Likelihood}
	\end{IEEEeqnarray}
	\par 
	The well-known optimization approaches could not be applied to obtain the maximum of the non-linear likelihood function of \eqref{eq:Likelihood} which contains the latent variables $M$ and $W$. Some approximate algorithms have been proposed to solve the hardship of optimization problems with latent variables in machine learning such as the expectation- maximization algorithm (EM)  \cite{murphy_machine_2012}, variational Inference  \cite{airoldi2008mixed} and block coordinate approach \cite{xu_block_2013}. 
	{We employ the \emph{Block Coordinate Ascent} algorithm to find the solution of the objective function in Eq \ref{eq:Likelihood}.} According to  \emph{Block Coordinate Ascent} approach, updating the parameters takes place in two main steps, \textbf{(i)} updating the first block, the membership function $M$, by fixing the second block, the parameters $\beta, I, W$, and  \textbf{(ii)}  updating the second block of parameters by fixing the first one. The log-likelihood function $\ell(\theta)$ is employed in our calculations that is more tractable than the likelihood in \eqref{eq:Likelihood} as,
	\begin{IEEEeqnarray}{rCl}
		\ell(\theta)&=& \sum_{(u,v)}a_{uv}\log\big(1-\exp(-M_u^T\beta M_v)\big) \IEEEnonumber \\
		&+& (1 - a_{uv})\mathcolor{black}{\big(-M_u^T \beta M_v\big)}\IEEEnonumber \\
		& +& \sum_{u}\sum_{k\in F}F_{uk}\log \frac{1}{1 + \exp(-\sum_{c_i}M_{uc_i}W_{kc_i})} \IEEEnonumber \\
		&+& (1 - F_{uk})\log \frac{\exp(-\sum_{c_i}M_{uc_i}W_{kc_i})}{1 + \exp(-\sum_{c_i}M_{uc_i}W_{kc_i})} 
		\label{eq:Loglik}
	\end{IEEEeqnarray}
	Details of the learning stage is follows.
	\subsection{Updating the parameters}
	The first step is to update the values of membership function $M_u$. To do so, it is required to derive the partial derivative of the log-likelihood function \eqref{eq:Loglik} with respect to $M_u$ as, 
	\begin{IEEEeqnarray}{rCl}
		\frac{\partial\ell(M_u)}{\partial M_u}  &=& \sum_{v\in N(u)} \beta M_v \frac{\exp(-M_u^T\beta M_v)}{1 - \exp(-M_u^T\beta M_v)}
		+\sum_{v \notin N(u)} -\beta M_v \label{eq:StdDeviationLL}\nonumber\\ 
		&+&\Big(F_{uk} - \frac{1}{1 + \exp(-\sum_{c_i}M_{uc_i}W_{kc_i})}\Big)W_{kc_i}
	\end{IEEEeqnarray}
	{where the set of neighbors of $u$ is represented by $N(u)$.} 
	{Each $M_u$ is first updated by gradient ascent in Eq~\eqref{eq:StdDeviationLL}, and then transformed to space of $[0, \infty)$ to meet the non-negativity property,
		\begin{equation}
		M_{u}(t+1) = max\big(0, M_{u}(t) + \alpha (\frac{\partial\ell(M_u)}{\partial M_u})\big) 
		\label{eq:MetaData:FinalUpdataLatent}
		\end{equation}
	}
	{where $\alpha$ is the learning weight parameter.}
	After updating  $M_u$, the second block of parameters $(I,W,\beta)$ are updated once at a time. \par 
	{First, the important probabilistic dependencies among the parameters are employed to simplify the calculations.}
	The conditional independence  between the parameters $(I,W,\beta)$, given $M_u$ is derived from the proposed probabilistic model. Also, the structure of the graphical model implies the probabilistic dependencies of $I$ to $\beta$ and $W$.  
	Taking into account the relation of  $M$ and $I$, and the chain rule to get  $\frac{\partial\ell}{\partial I}$ for a node $u$ as,
	\begin{equation}
	\frac{\partial\ell}{\partial I} = \frac{\partial\ell(M_u)}{\partial M_u} \times \frac{\partial M_u}{\partial I}
	\label{eq:GeneralFormOfI}
	\end{equation}
	The $\frac{\partial M_u}{\partial I}$ is obtained from Eq. \eqref{eq:ProposedMembership} by, 
	\begin{equation}
	\frac{\partial M_u}{\partial I} = S_u \times \frac{\exp (-\sum_{k\in I} S_{uk}.I_{c_ik} )}{[1 + \exp (-\sum_{k\in I}S_{uk}.I_{c_ik} )]^2}
	\label{eq:DerivativeMResI}
	\end{equation}
	Eq. \eqref{eq:DerivativeMResI} and Eq. \eqref{eq:StdDeviationLL} imply the updating procedure of $I$ as, 
	\begin{equation}
	I(t + 1) = I(t) + \alpha (\sum_{u \in V}\frac{\partial\ell(M_u)}{\partial M_u} \times \frac{\partial M_u}{\partial I})
	\label{eq:FinalFormI}
	\end{equation}

	In the next step, parameter $W$, which is responsible for the correlation levels between generative features and the community is updated according to the following,
	\begin{equation}
	\frac{\partial\mathcal\ell}{\partial W_{kc_i}} = \sum_{u}(F_{uk} - \frac{1}{1 + \exp(-\sum_{c_i}M_{uc_i}W_{kc_i})})M_{uc_i}
	\label{eq:FindDerivationW}
	\end{equation}
	In a similar way, the final updating form of parameter $W$ takes the form as the following,
	\begin{equation}
	W_{kc_i}(t+1) = W_{kc_i}(t) + \alpha(\frac{\partial\ell}{\partial W_{kc_i}})
	\label{eq:MetaData:Updata:W}
	\end{equation}

	To update $\beta$, the derivation of the log-likelihood function with respect to $\beta$ is calculated as,
	\begin{IEEEeqnarray}{rCl}
		\frac{\partial\ell}{\partial \beta} &=& \sum_{v\in N(u)} -M_u^TM_v\times \frac{\exp(-M_u^T\beta M_v)}{1 - \exp(-M_u^T\beta M_v)}  \IEEEnonumber \\
		& +& \sum_{v \notin N(u)} -M_u^TM_v   \label{eq:FindDerivationBeta}
	\end{IEEEeqnarray}
	The Eq. \eqref{eq:FindDerivationBeta} provides the updating procedure for $\beta$ as,
	\begin{equation}
	\beta(t + 1) = \beta(t) + \alpha (\frac{\partial\ell(M_u)}{\partial \beta}) 
	\label{eq:MetaData:Updata:Beta}
	\end{equation}
	
	
	\begin{algorithm}[h]
		\algsetup{linenosize=\tiny}
		\scriptsize
		\caption{\small{Probabilistic Feature based Community Detection(PFCD)}}
		\label{alg:MetaData}
		\begin{algorithmic}[1]
			\STATE {\bfseries Input:}  $G = (V,E), Features, Number\ of\ communities(k)$.
			\STATE {\bfseries Output:}  $M$ Community memberships of each node.
			\STATE {\bfseries Initialize:} Initializing parameters. 
			\STATE {$t \gets 0$}
			\WHILE {$|M_u{(t+1)}-M_u{(t)}|<= threshold$}
			\STATE {$t \gets t + 1$}
			\FOR {$i = 1$ \TO $|V|$}
			\STATE {$DevM= findDerivationM()$ }
			\STATE {\bfseries Update:}  $M_{v_i}(t+1)= UpdataM(DevM,M_{v_i}(t))$.
			\ENDFOR
			\FOR {$i = 1$ \TO $|S|$}
			\STATE {$DevI= findDerivationI()$ }
			\STATE {$I(t+1) = UpdateI(I(t),DevI)$ }
			\ENDFOR
			\FOR {$i = 1$ \TO $|F|$}
			\STATE {$DevW= findDerivationW()$ } 
			\STATE {$W(t+1) = UpdateW(W(t),DevW)$ }
			\ENDFOR
			\FOR {$i = 1$ \TO $k$}
			\STATE {$Dev\beta= findDerivation\beta()$ }
			\STATE {$\beta(t+1) = UpdateBeta(\beta(t),Dev\beta)$ }
			\ENDFOR
			\ENDWHILE			
		\end{algorithmic}
	\end{algorithm}
	
	\subsection{PFCD algorithm}
	The proposed approach, \emph{PFCD}  (\textbf{P}robabilistic \textbf{F}eature based \textbf{C}ommunity \textbf{D}etection) is represented in Algorithm \ref{alg:MetaData}. According to \emph{PFCD}, the inputs are the network structure ($G$), the features ($S \cup F$), the number of assortative features ($|S|$), the number of generative features ($|F|$) and the number of communities ($k$). The final output is the membership function of each node. The relationship  between different types of features and communities can also be two important output of the proposed algorithm with the aim of interpretation. After initialization of the parameters, the main part of the algorithm is performed in an iterative manner. The algorithm stops when the absolute difference between the subsequent log-likelihoods of the model (Eq. \eqref{eq:Likelihood} ) is less than a threshold parameter, which is set to $0.001$ in our setting. 
	Function \emph{findDerivationM} is applied to compute the derivation of the \emph{log-likelihood} function with respect to $M$ based on  Eq. \eqref{eq:StdDeviationLL}.
	The update procedure on $M$ is performed by \emph{UpdataM} based on Eq. \eqref{eq:MetaData:FinalUpdataLatent}. 
	When updating the membership function for all nodes is finished, the next step is to update the parameters $I$, $W$ and $\beta$. The \emph{findDerivationI} calculates the derivation of the \emph{log-likelihood} function with respect to $I$ by Eq. \eqref{eq:DerivativeMResI} and function \emph{UpdateI} updates  $I$ by Eq. \eqref{eq:FinalFormI}. $W$ which captures the correlation level between the communities and the generative features, is updated by \emph{FindDerivationW} and \emph{UpdateW} according to equations \eqref{eq:FindDerivationW} and \eqref{eq:MetaData:Updata:W}. 
	$\beta$ is updated based on equations \eqref{eq:FindDerivationBeta} and \eqref{eq:MetaData:Updata:Beta}. The update procedure is repeated until the convergence criterion is met.
	\subsection{Computational Complexity}
	The complexity of PFCD in each iteration is linearly dependent on the number of communities, assortative and generative features in network. The process of updating of \emph{PFCD} consists of two parts, updating the membership value of each node toward each community and updating the weight of each feature to each community. The membership value is updated based on Eq.s \eqref{eq:StdDeviationLL} and \eqref{eq:MetaData:FinalUpdataLatent}.  For each given node $u$, the process considers the membership of its neighbors $M_v , v\in N(u)$ and non-neighbors towards communities $M_v, v \notin N(u)$. Therefore, for each given node Eq. \ref{eq:StdDeviationLL} takes $O(|N(u)|(k^2 +k))\sim O(|N(u)|k^2)$  (because of multiplying $M_u^T\beta M_v$) time for the neighbors and $O(|N(u)|k^2)$ for the non-neighbors. After iterating on all nodes the time complexity  is $O(|E|\times k^2)$.
	
	
	The time complexity for each given node of the second component of Eq. \ref{eq:StdDeviationLL} is related to $O(k\times |F|)$. As a result the time complexity of the second component iterating on all nodes would  end up with $O(|V|\times k \times |F|)$. 
	The next step is the weight computation for each feature, \emph{\{I,W\}}, and the community interaction matrix, \emph{$\beta$}. 
	According to Eq. \ref{eq:DerivativeMResI}, the time complexity is related to multiplying $S_u\times I_{c_i}$ which can be done in $O(|V|\times k \times |S|)$. Accordingly, 
	updating the parameter $W$ has complexity of $O(|V|\times k \times |F|)$ and the matrix \emph{$\beta$} takes $O(|E|\times k^2)$ for updating its values by considering Eq. \eqref{eq:MetaData:Updata:Beta}. 
	Therefore, the proposed method would have the complexity of $max(O(|E|\times k^2), O(|V|\times k \times |F|))$. }

\section{Experiments}
\label{sec5}
Synthetic and real-world networks are used to evaluate performance of  \emph{PFCD}.  The efficiency of the proposed model is demonstrated on the state-of-the-art community detection methods by considering the feature based methods and structure based ones'.  On state-of-the-art feature based methods,  \emph{Cesna} \cite{yang2013community}, \emph{JCDC} \cite{zhang_community_2016}, \emph{NC} \cite{newman2016structure}, \emph{BAGC} \cite{xu2012model}, \emph{SDP} \cite{yan2016convex}, and \emph{CASC} \cite{binkiewicz_covariate-assisted_2017} are employed in the experiments. On state-of-the-art structure based methods, \emph{BigClam} \cite{yang2013overlapping}, \emph{Fast-Greedy} \cite{gregory_finding_2010}, \emph{Infomap} \cite{rosvall_maps_2008}, \emph{Louvain} \cite{blondel2008fast}, and \emph{COMBO} \cite{sobolevsky_general_2014} are applied in the experiments. Table\ref{tbl:GeneralMethod} summarizes these algorithms.\par
\begin{table}[t!]
	\centering
	\small
	\caption{Overview of the state-of-the-art algorithms}
	\resizebox{.5\textwidth}{!}{
		\begin{tabular}{llc}
			\toprule[1.5pt]
			Methods & Description & Reference\\ 
			\midrule	
			\multicolumn{3}{c}{Feature based Community Detection Methods}\\
			\midrule			
			Cesna &  Feature enabled Generative model  & \cite{yang2013community}\\
			JCDC& Joint Feature based Community Detection Criterion& \cite{zhang_community_2016}\\
			NC& Modified Feature based Stochastic Block Model& \cite{newman2016structure}\\
			BAGC & Bayesian Graph Clustering &\cite{xu2012model}\\
			SDP &Semi-Definite Programming&\cite{yan2016convex}\\
			CASC& Covariance Assisted Spectral Clustering & \cite{binkiewicz_covariate-assisted_2017}\\
			\midrule
			\multicolumn{3}{c}{Structure based Community Detection Methods}\\
			\midrule 			
			BigClam & Find Overlapping community & \cite{yang2013overlapping}\\
			Fast-Greedy & Fast Overlapping community detection & \cite{gregory_finding_2010} \\ 
			Infomap &  Find Overlapping community by probabilistic flows &\cite{rosvall_maps_2008}\\
			Louvain & Heuristic method for detecting non-overlapping community & \cite{blondel2008fast}\\
			COMBO & Find overlapping and non-overlapping communities & \cite{sobolevsky_general_2014}\\
			\bottomrule[1.5pt]
		\end{tabular} 
	}
	\label{tbl:GeneralMethod}
\end{table}  
Initially, synthetic networks are used to examine the proposed approach. Then, we compare the performance of \emph{PFCD} on benchmark real networks with ground-truth communities.  {In our experiments, the ground-truth communities and presumed number of communities are used for all of the methods to yield a fair comparison as highly recommended \cite{fortunato2016community}}.   {Experiments are run on a desktop PC with 4GB memory and Core i5 CPU under JAVA using JGraphT library. The source codes are provided in the supplementary information. The weight parameter $\alpha$ is set as $0.001$ in the experiments. The $\beta$ is initialized based on the conductance measure approach \cite{Gleich2012}. The algorithms are run with their default parameters.}
\subsection{Evaluation criteria}
\label{EvalCriteria}
Two well-known evaluation criteria, the \emph{F1 score} and the \emph{NMI}, are applied to measure the accuracy of the community detection algorithms as compared to the ground-truth communities \cite{fortunato2016community}.  \emph{F1 score} is a standard evaluation measure in machine learning and  community detection tasks, which quantifies the relative frequency of the number of correct detections of the members in each community based on the gold-standard information. The second performance measure is \emph{NMI} which is the mutual information of the similarity (or dissimilarity) between the discovered communities and the ground--truth ones'.
\subsection{Synthetic networks}
Synthetic networks are generated based on the degree-corrected stochastic block model \cite{zhang_community_2016}. The generation process is performed at two phases. At the first phase, the structure of network is shaped and the features are provided to each node at the second phase.

At the first phase, a pair of nodes $(i,j)$ are sharing an edge independently  from the other pairs. The probability of an edge generation between any pair of nodes depends on whether they are in a same community or not. If they share a community, the probability would be $\beta\theta_i \theta_j$, otherwise is $r\theta_i\theta_j\beta$. Parameter $\beta$ indicates the level of interaction between any pair of communities, such that higher level of  $\beta$ for a pair of communities $(i,j)$ results in more interactions of them. The $r$ is used for handling the density inside the community and parameters $\theta_i, \theta_j$ are used for controlling the degree of nodes. To avoid homogeneity in the generated networks, we set 10\% of nodes inside a community as hub by setting  $\theta_i = 10$ and for non-hub nodes $\theta_i = 1$. 
We set $\beta=0.1$ and $r = 0.25$ in generating the networks. The average degree of the resulted network is around $31$. 
After shaping the structure of network, at the second phase we generate features for two communities from the Gaussian distribution $\mathcal{N}(\mu,1)$, for nodes of the first community and $\mathcal{N}(-\mu , 1)$  for nodes of the second community. As $\mu$ increases, the feature of each community becomes stronger. To reveal the impact of nodal features on communities, three different networks are generated with $N=\{1000, 2000, 5000\}$ by considering three different scenarios $\mu = \{2, 3, 5\}$ for each network. 
\textcolor{black}{A summary of properties of the synthetic networks are given in Table \ref{tbl:SyntheticNetworks}.}
\begin{table}[t!]
	\centering
	\caption{Main properties of the synthetic networks.}
	\label{tbl:SyntheticNetworks}
	\begin{tabular}{lccc}
		\toprule[1.5pt]
		Network & Nodes& Edges  & Communities \\ 
		\midrule 
		\textit{1000-2}  & 1000 & 51578 & 2  \\
		\textit{1000-3.5} & 1000& 56401 & 2 \\
		\textit{1000-5}  & 1000 & 52488 &  2 \\		
		\textit{2000-2}  & 2000 & 177719& 2 \\
		\textit{2000-3.5}  & 2000 & 181288 & 2\\			
		\textit{2000-5}  & 2000 & 177592  & 2 \\
		\textit{5000-2}  & 5000 & 819220 & 2 \\
		\textit{5000-3.5}  & 5000 & 783694 & 2 \\
		\textit{5000-5}  & 5000 & 845994 & 2  \\
		\bottomrule[1.5pt]    
	\end{tabular} 
\end{table}

\begin{figure}[H]
	\centering
	\includegraphics[scale=0.5]{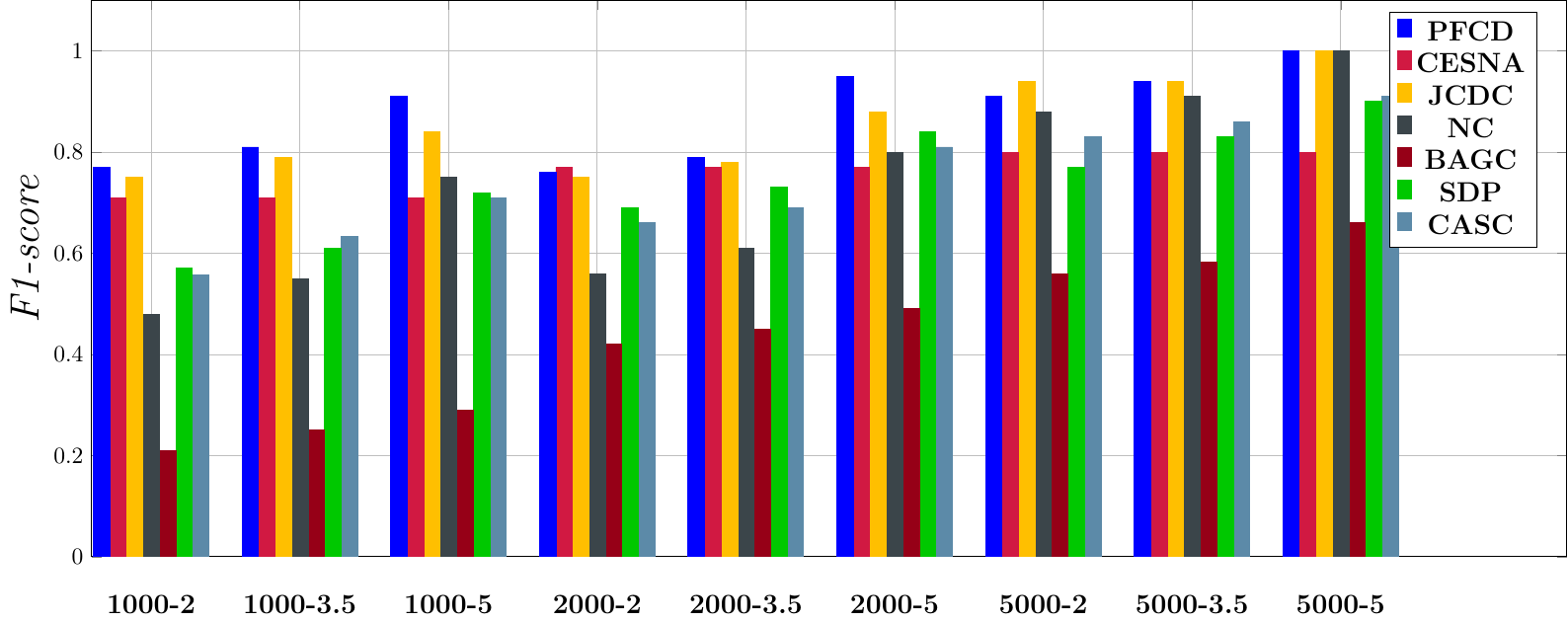}
	\caption{\label{fig:SimResFScore}	\textcolor{black}{Results of \emph{PFCD} compared with feature based community detection methods (Table \ref{tbl:GeneralMethod}) in terms of F1-Score, where the horizontal axis represents three different scenarios for each generated network with sample sizes from $1000$ to $5000$.}}
\end{figure}

\begin{figure}[H]
	\centering
	\includegraphics[scale=0.5]{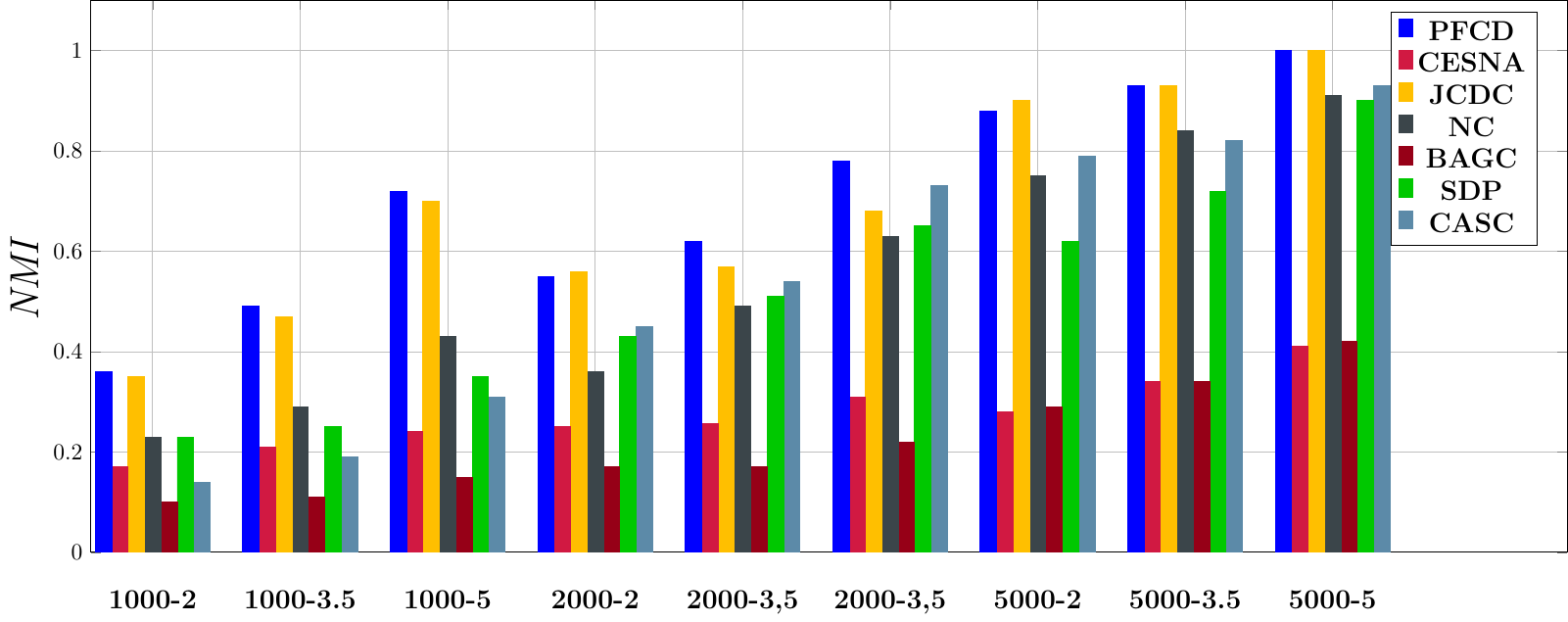}
	\caption{\label{fig:SimResNMI}	\textcolor{black}{Results of \emph{PFCD} compared with the feature based community detection methods (Table \ref{tbl:GeneralMethod}) in terms of NMI, where the horizontal axis presents three different scenarios for each generated network with sample sizes from $1000$ to $5000$.}}
\end{figure}

\begin{figure}[h]
	\centering
	\includegraphics[scale=0.8]{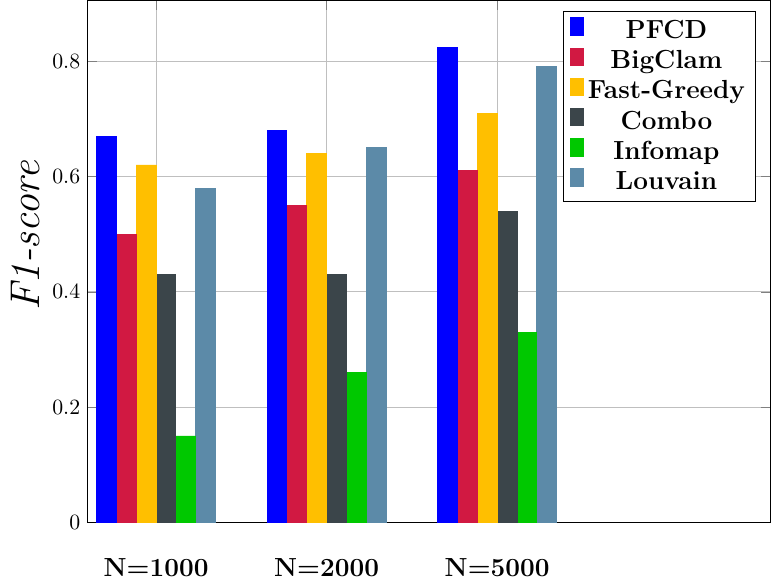}
	\caption{\label{fig:SimResFMeasStruct}		\textcolor{black}{Results of \emph{PFCD} compared with the structure based community detection methods (Table \ref{tbl:GeneralMethod}) in terms of  F1-Score, where the horizontal axis presents three generated networks with sample sizes from $1000$ to $5000$.}}
\end{figure}

\begin{figure}[h]
	\centering
	\includegraphics[scale=0.82]{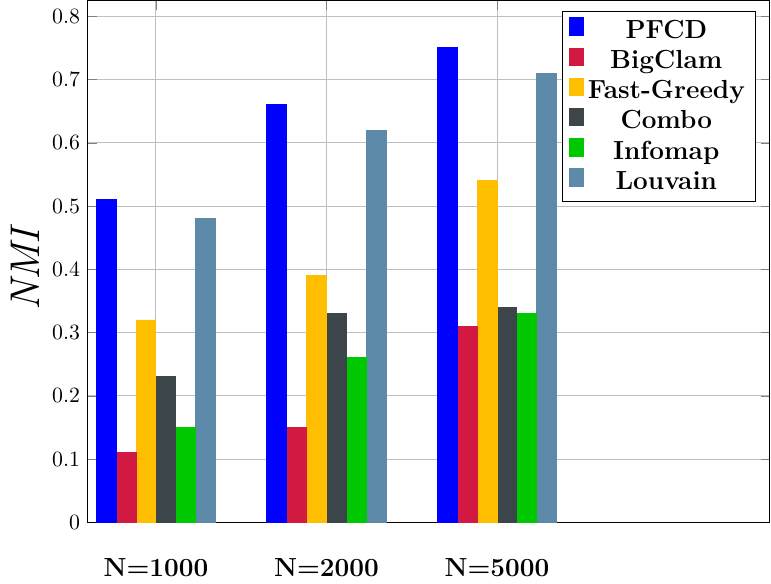}
	\caption{\label{fig:SimResNMIStruct}	\textcolor{black}{Results of \emph{PFCD} compared with the structure based community detection methods  (Table \ref{tbl:GeneralMethod}) in terms of NMI, where the horizontal axis presents three generated networks with sample sizes from $1000$ to $5000$.}}
\end{figure}

The performance of \emph{PFCD} is demonstrated on these generated networks by using the state-of-the-art feature based methods in Table \ref{tbl:GeneralMethod}.
The results are shown in Figures \ref{fig:SimResFScore} and \ref{fig:SimResNMI}. As $\mu$ increases, the influence of features on community structures becomes stronger. 
\textcolor{black}{Due to the number of features in the generation of networks, there exists a slight difference between the proposed approach and other feature-enabled community detection methods. In addition, the main aim of the experiments on synthetic networks is to demonstrate the impact of features on the performance of community detection (Figures \ref{fig:SimResFMeasStruct} and \ref{fig:SimResNMIStruct}).}  We compare the proposed approach with the well-known structure-based methods in Table \ref{tbl:GeneralMethod} to reveal the importance of features on community detection process by using  F1-Score and NMI criteria.  Figures \ref{fig:SimResFMeasStruct} and \ref{fig:SimResNMIStruct} indicate that the proposed method outperforms the well-known structure based methods. 


\subsection{Real networks}
We examine our approach on a number of benchmark real-world network dataset. The networks are from different domains including  economy, biology, ecology, and social.

\subsubsection{Dataset description}
Three social friendship networks, namely \emph{Lawyer}, \emph{CalTech}, and \emph{Rice}, are used in this study. The \emph{Lawyer} dataset is derived from a study of corporate law partnership that was developed  in a Northeastern US corporate law firm. It contains the friendship networks among 71 attorneys (partners and associates) of this company. There are several features for the members as part of the dataset, such as seniority, formal status, working office location, gender, law school attended,  working hours, the years of activity, and attitudes on various management  options \cite{lazega2001collegial}. 
We consider two Facebook subnetworks, namely \emph{CalTech} and \emph{Rice}, from the Facebook-100 dataset, which consists of the Facebook networks for 100 colleges and universities in the US.  There are links between the members (student or faculty) inside each school, and also nodal features including the status of student/faculty, major, senior or junior, dormitory, year and high school information \cite{traud2012social}. \par
Five information networks are used in our experiments, including \emph{DBLP}, \emph{ArXiv},  \emph{WTrade}, \emph{Internet}, and \emph{PolBlogs}.
In \emph{DBLP} repository, the nodes and edges represent the authors and co-authorship relationships. 20 keywords are extracted from the title of papers to represent four different fields: Data-Mining, Computer Graphics, Artificial Intelligence and Databases. The keywords include ``classification'', ``cluster'', ``graphic'' and ``human''. 
In \emph{ArXiv}, the nodes represent papers and the edges show citations between them. The features denote how often a specific keyword appears in the abstract of a paper. The \emph{ArXiv} network contains 30 distinct keywords.  \emph{WTrade}  is a dataset of various manufactures from 80 countries on metal trade commodities in 1994. The edges show the exports from one country to another for metal commodities. The nodes are countries with features such as the continent, position in world system and GDP \cite{de2011exploratory}. \emph{Internet} network is a topological network where each node represents Autonomous Systems (AS) and the edges represent the path from an AS to another one. Communities are countries of the registered AS. 
\emph{PolBlogs} is a network of hyperlinks between weblogs on US politics, recorded in 2005 by Adamic and Glance  \cite{adamic2005political}. Each node is represented by its political affiliation, conservative or liberal.
{\emph{Patent} \cite{gunnemann2013spectral}  is a large citation networks on the utility patents granted which is maintained by the National Bureau of Economic Research during January 1, 1963 to December 30, 1999.}
\par 
Two biological networks \emph{Predator-Prey} and  \emph{Malaria} are employed in the experimental settings.
\emph{Predator-Prey} is an ecological network about 488 marine creatures living in the WeddellSea. Each creature has different features such as feeding type, feeding mode, environment and body mass \cite{jacob2011role}. 
\emph{Malaria} dataset is a biological network of genetic sequences from the malaria parasites \cite{larremore2013network,rask2010plasmodium}. In this network, the nodes represent  297 genes and their various shared amino acid substrings. 
The common process of  recombination among genes to produce proteins generates a natural two-mode network which consists two types of nodes: genes in HVRs (highly variables regions) and every HVR  has different set of edges among the same nodes \cite{rask2010plasmodium}. 
A summary of the datasets are given in Table  \ref{tbl:Networks}. 
\begin{table}[t!]
	\centering
	\caption{Summary of the real network datasets.}
	\label{tbl:Networks}
	\resizebox{0.5\textwidth}{!}{
		\begin{tabular}{lcccccc}
			\toprule[1.5pt]
			Network & Nodes& Edges& Domain & Features &Communities \\ 
			\midrule 
			\textit{Lawyer} \cite{lazega2001collegial} & 71 & 379 & Social & Status, Office, Years & 2  \\
			\textit{CalTech} \cite{traud2012social} & 769 & 16656 & Social & Gender, Class year, Major, Residence & 9 \\
			\textit{Rice} \cite{traud2012social} & 4087 & 184828 &  Social & Gender, Class year, Major, Residence & 9 \\		
			\textit{DBLP} \cite{gunnemann2013spectral} & 774 & 1757 & Social & Extracted Keywords from papers & 20 \\
			\textit{PolBlogs} \cite{adamic2005political} & 1490 & 19090 & Political & Political Affiliation & 2\\			
			\textit{World Trade} \cite{de2011exploratory} & 80 & 876 & Economical & Continent, Positions & 6 \\
			\textit{Malaria} \cite{larremore2013network,rask2010plasmodium} & 307 & 7759 & Biological & Cys-PoLV labels & 4 \\
			\textit{WeddellSea} \cite{jacob2011role} & 488 & 15435 & Ecological & Feeding type and mode, Body mass, Environment & 4  \\
			\textit{ArXiv} \cite{gunnemann2013spectral} & 856 & 2660 & Scientific & Abstract Keywords of papers & 9 \\		
			\textit{Internet} \cite{hric2014community} & 46676 & 262953 & Technological & Country & 131\\			
			\textit{Patent} \cite{gunnemann2013spectral} & 100000 & 188631 & Citation & Year, Country, PatentClass, Assigned Code & 15   \\	
			\bottomrule[1.5pt]    
		\end{tabular} 
	}
	
\end{table}
\subsubsection{Experimental results}
\begin{figure}[h]
	\centering
	\includegraphics[scale = 0.57]{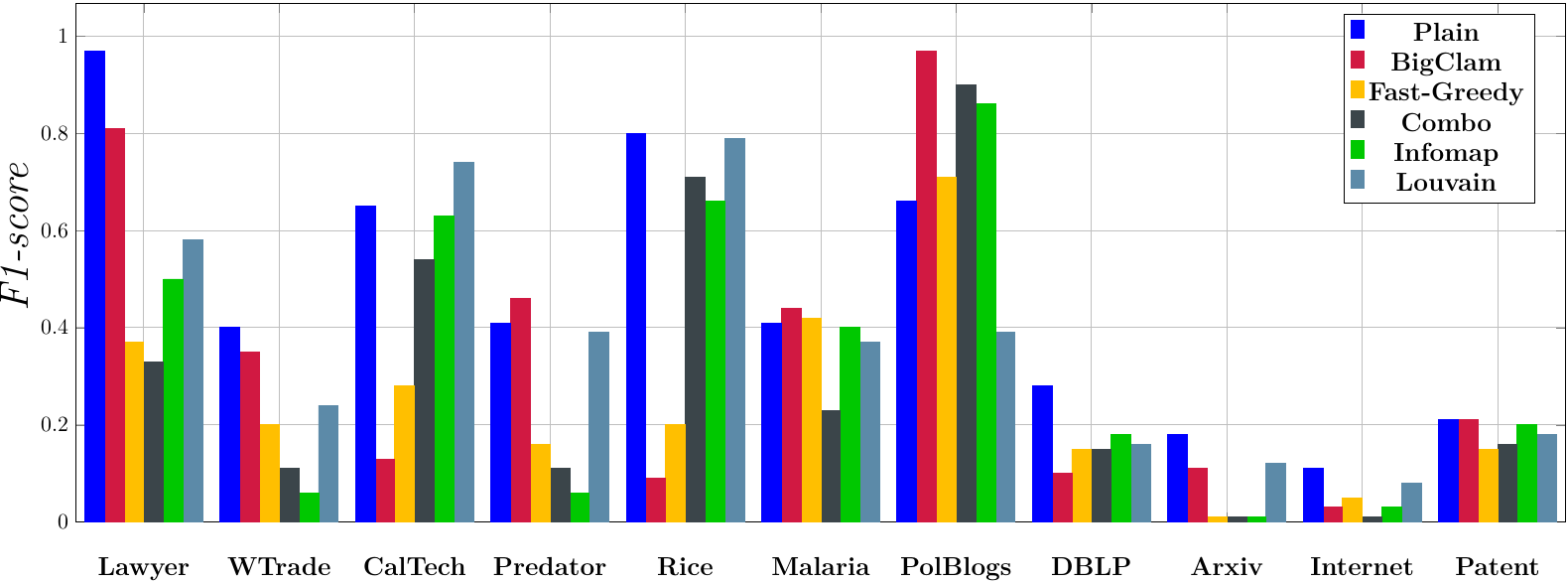}
	\caption{	\textcolor{black}{Results of the proposed method compared with others by just considering connectivity structure, \emph{Plain}, in terms of F1-score. The baseline methods include \emph{BigClam} \cite{yang2013overlapping}, \emph{Fast-Greedy} \cite{gregory_finding_2010}, \emph{Infomap} \cite{rosvall_maps_2008}, \emph{Louvain} \cite{blondel2008fast}, and \emph{COMBO} \cite{sobolevsky_general_2014} } }
	\label{fig:Fscore}
\end{figure}

\begin{figure}[h]
	\centering
	\includegraphics[scale = 0.57]{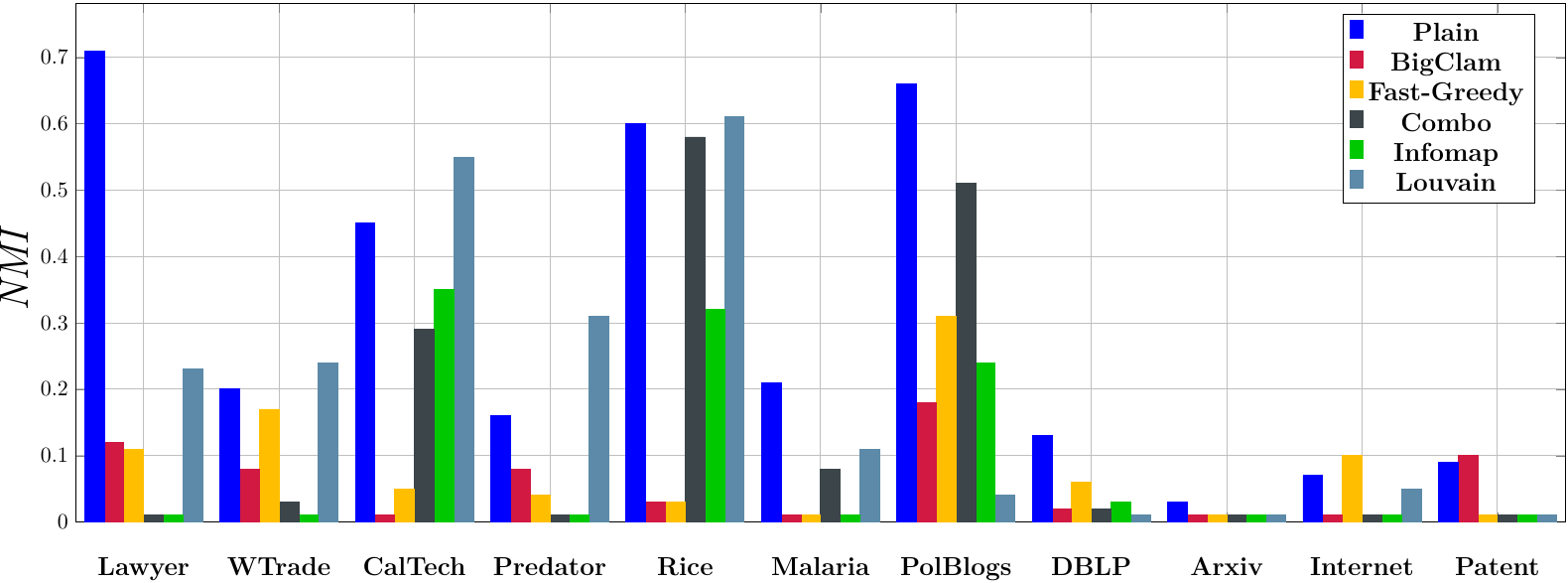}
	\caption{\textcolor{black}{Results of the proposed method compared with others by just considering connectivity structure, \emph{Plain}, in terms of NMI. The baseline methods include \emph{BigClam} \cite{yang2013overlapping}, \emph{Fast-Greedy} \cite{gregory_finding_2010}, \emph{Infomap} \cite{rosvall_maps_2008}, \emph{Louvain} \cite{blondel2008fast}, and \emph{COMBO} \cite{sobolevsky_general_2014}.} }
	\label{fig:nmi}
\end{figure}

\begin{figure*}[!t]
	\centering
	\begin{minipage}{0.9\textwidth}
		\centering
		\includegraphics[width=0.9\linewidth]{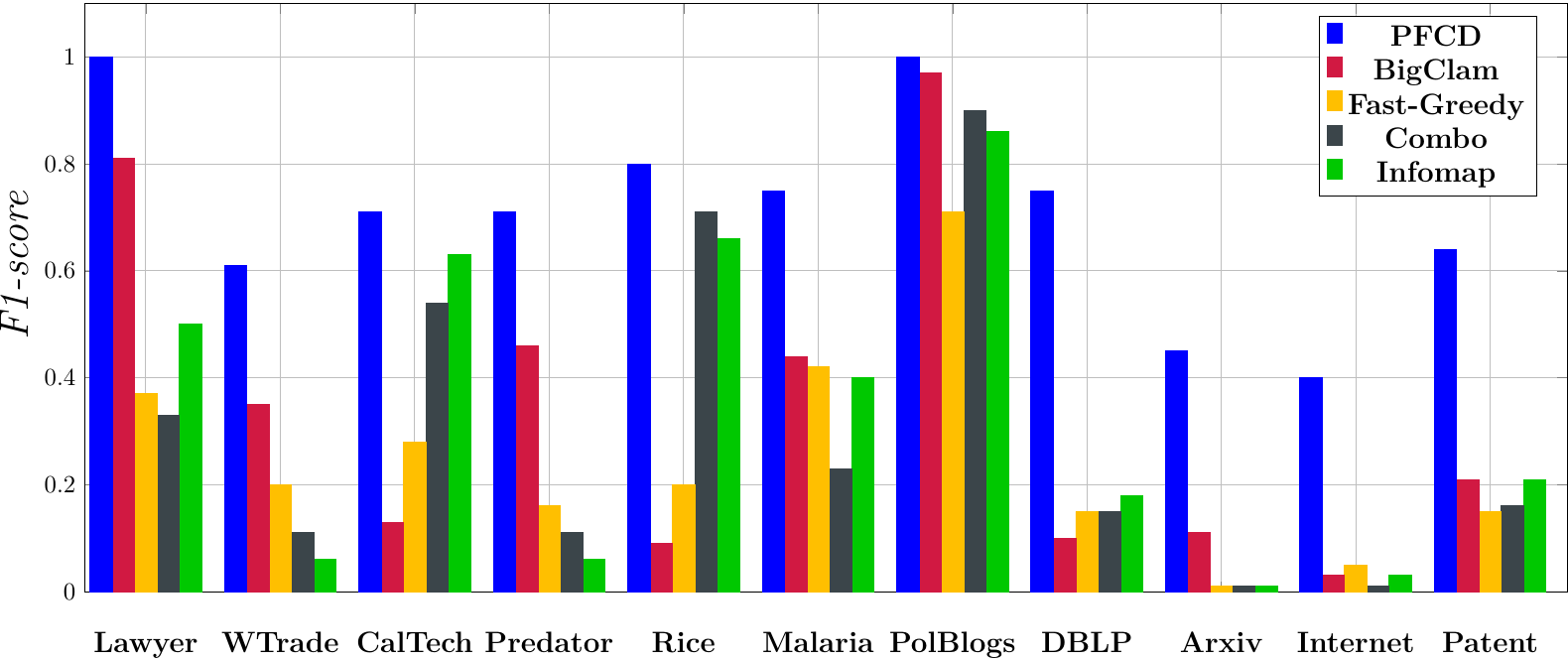}
		\subcaption{}
		\label{SubFig:Weddell}
	\end{minipage}%
	
	\begin{minipage}{0.9\textwidth}
		\centering
		\includegraphics[width=0.9\linewidth]{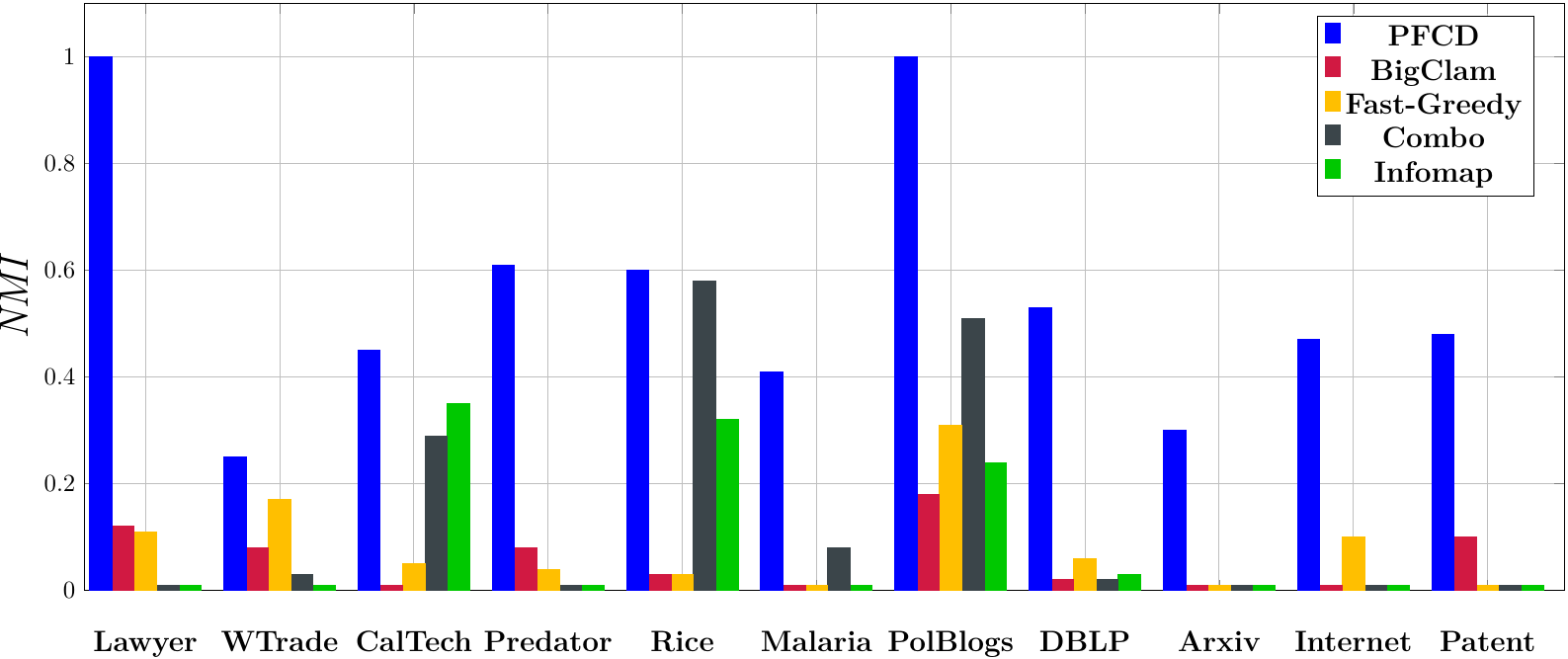}
		\subcaption{}
		\label{SubFig:WorldTrade}
	\end{minipage}
	\caption{\label{fig:withoutMetaDataEvaluation_nmi} Results of \emph{PFCD} compared with the well-known structure based methods. The baseline methods include  \emph{BigClam} \cite{yang2013overlapping}, \emph{Fast-Greedy} \cite{gregory_finding_2010}, \emph{Infomap} \cite{rosvall_maps_2008}, \emph{Louvain} \cite{blondel2008fast}, and \emph{COMBO} \cite{sobolevsky_general_2014} (Table \ref{tbl:GeneralMethod})}
\end{figure*}

At first, we consider the performance of \emph{PFCD} without taking into account of features, \emph{Plain}, along with some of the state-of-the-art structure based methods such as, \emph{BigClam} \cite{yang2013overlapping}, \emph{Fast-Greedy} \cite{gregory_finding_2010}, \emph{Infomap} \cite{rosvall_maps_2008}, \emph{Louvain} \cite{blondel2008fast}, and \emph{COMBO} \cite{sobolevsky_general_2014} (see Table\ref{tbl:GeneralMethod}).
\textcolor{black}{An automatic strategy is used on threshold specification in \emph{Plain} approach, where each node is assigned to the community with the greater membership value than the average of the memberships of all nodes \cite{fortunato2016community}. }
Figures \ref{fig:Fscore} and \ref{fig:nmi} show the results. It reveals that the proposed method is able to perform as well as the other structure-based methods. Moreover, its result on some networks like Lawyer, CalTech and Rice is so competitive compared with results of its original version. Actually, the PFCD algorithm without nodal features considers community as dense sub-graphs like most of other structure-based methods.
\textcolor{black}{The obtained results by considering just the structural properties are not good enough (less than 0.2) on certain type of networks such as DBLP, Arxiv, Internet, and Patent due \textcolor{black}{to the fact that} the community structures consist of  the dense sub-graphs and assortative modules \cite{gunnemann2013spectral}. In addition, structure based methods perform better than the \emph{Plain} on some networks such as \emph{Predator} and \emph{PolBlogs} (Figure \ref{fig:Fscore}) due to having small number of features. These results reveal the impact of features from the viewpoint of types and numbers on the detection of community structures in real networks, where Figure \ref{fig:withoutMetaDataEvaluation_nmi} demonstrate the usefulness of features on the community detection process. }  

Then, the \emph{PFCD} is compared with the the well-known structure based methods in Table \ref{tbl:GeneralMethod}. Figure \ref{fig:withoutMetaDataEvaluation_nmi} represents the results in terms of  \emph{F1-score} and \emph{NMI}. We observe that considering nodal features in the community detection process leads to the superiority of  \emph{PFCD} compared to the algorithms that are only based on structural information. 
The difference is specially well pronounced in some networks, such as  \emph{Lawyer}, \emph{Malaria}, \emph{Predator-Prey}, \emph{DBLP}, \emph{ArXiv} and \emph{Internet}, for which \emph{PFCD} has higher \emph{F1-score} and \emph{NMI} than other algorithms.

\begin{figure*}[!t]
	\centering
	\begin{minipage}{0.9\textwidth}
		\centering
		\includegraphics[width=0.9\linewidth]{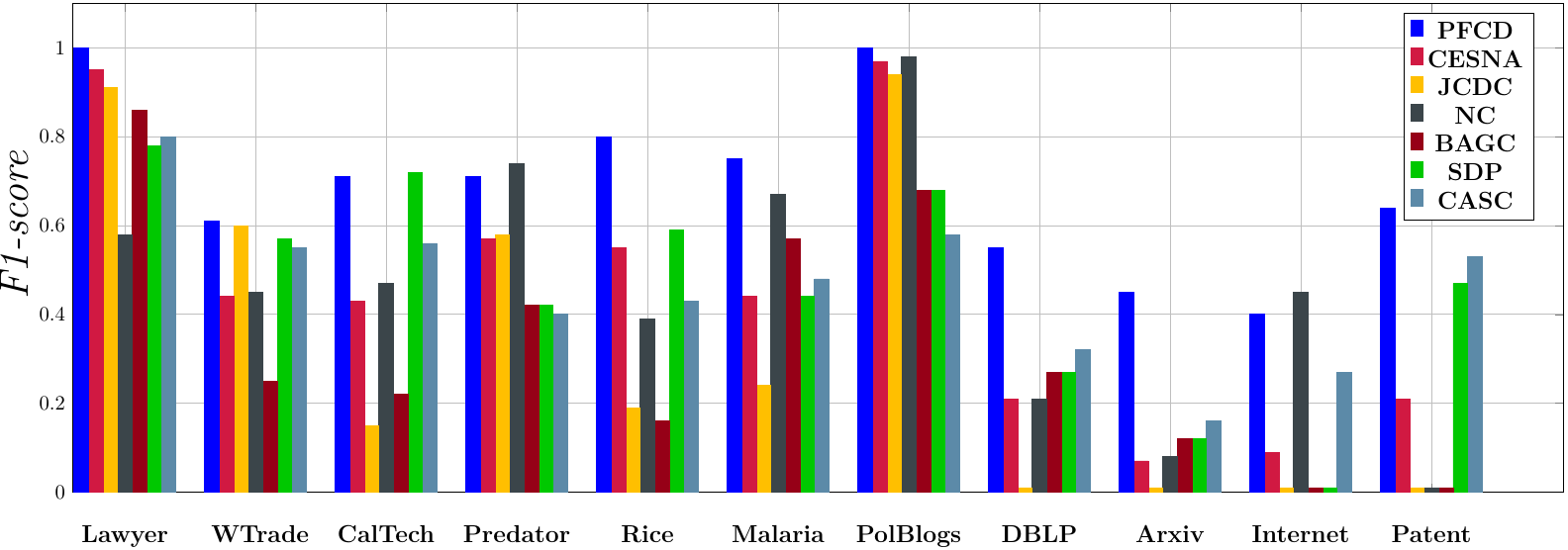}
		\subcaption{}
		\label{SubFig:Weddell}
	\end{minipage}%
	
	\begin{minipage}{0.9\textwidth}
		\centering
		\includegraphics[width=0.9\linewidth]{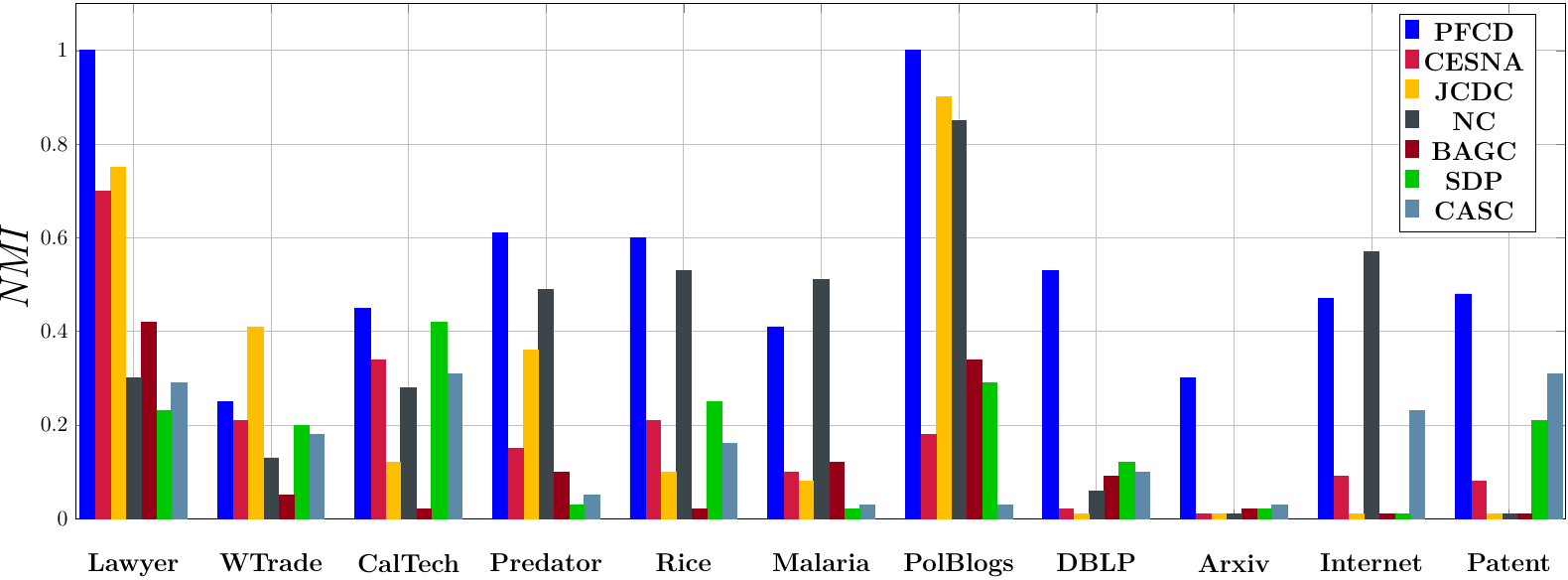}
		\subcaption{}
		\label{SubFig:WorldTrade}
	\end{minipage}
	\caption{\label{fig:MetaDataEvaluation} Results of \emph{PFCD} compared with the benchmark feature based methods in terms of F1--score and NMI. The baseline methods include \emph{Cesna} \cite{yang2013community}, \emph{JCDC} \cite{zhang_community_2016}, \emph{NC} \cite{newman2016structure}, \emph{BAGC} \cite{xu2012model}, \emph{SDP} \cite{yan2016convex} and \emph{CASC} \cite{binkiewicz_covariate-assisted_2017} (Table \ref{tbl:GeneralMethod})}
\end{figure*}

Moreover, the performance of our approach are demonstrated with the state-of-the-art feature based algorithms in Table\ref{tbl:GeneralMethod}. The obtained results are shown in Figure \ref{fig:MetaDataEvaluation}. We observe that \emph{PFCD}  outperforms others in almost all experiments. Specifically in the networks for which features are completely dependent on the community structures, such as \emph{WTrade}, \emph{PolBlogs}, \emph{DBLP} and \emph{ArXiv} networks, the proposed approach performs better than the others. The results show that higher dependency between features and community structure can lead to higher accuracy in the community detection process. 
While  \emph{JCDC} performs well on small networks, it fails to accurately detect communities in large networks due to its dependency on multiple tuning parameters. \emph{NC} considers only one type of feature as a metadata and fails to precisely detect communities. The results return similar weak performance of  \emph{CESNA}, on  networks such as \emph{WTrade}, \emph{Predator-Prey}, \emph{Malaria}, \emph{PolBlogs} and \emph{ArXiv}.
{On the division of the features into assortative  and generative categories, we used NMI to select the appropriate assortative features due to the high level impact on the community structures. After selection the assortative features, the remaining features are used in the category of generative features. 
	The details are reported in Table \ref{tbl:assortativeNetworks}.}
\begin{table}[h]
	\centering
	\caption{The properties of features in the real network datasets.}
	\label{tbl:assortativeNetworks}
	\resizebox{0.5\textwidth}{!}{
		\begin{tabular}{lcc}
			\toprule[1.5pt]
			Network & Features& Assortative Feature\\ 
			\midrule 
			\textit{Lawyer} \cite{lazega2001collegial} & Status, Office, Years & Status \\
			\textit{CalTech} \cite{traud2012social} &Gender, Class year, Major, Residence & Class Year \\
			\textit{Rice} \cite{traud2012social}  & Gender, Class year, Major, Residence  & Class Year\\		
			\textit{DBLP} \cite{gunnemann2013spectral} & extracted keywords from papers & Keyword  \\
			\textit{PolBlogs} \cite{adamic2005political}  & Political Affiliation(Liberal, Conservative) & Political Affiliation \\			
			\textit{World Trade} \cite{de2011exploratory}& Continent, Positions & Continent \\
			\textit{Malaria} \cite{larremore2013network,rask2010plasmodium}  & Cys-PoLV labels & Cys-PolV \\
			\textit{WeddellSea} \cite{jacob2011role} & Feeding type, Feeding mode, Body mass, Environment & Feeding Type \\
			\textit{ArXiv} \cite{gunnemann2013spectral} & Keywords from abstract of papers & Keywords \\		
			\textit{Internet} \cite{hric2014community} & Country & Country \\	
			\textit{Patent} \cite{gunnemann2013spectral} & Year, Country, PatentClass, Assigned Code & PatentClass   \\
			\bottomrule[1.5pt]    
		\end{tabular} 
	}
\end{table}			

\begin{table}[h]
	\caption{\label{tbl:execution} {The execution running time of \emph{PFCD} along with \emph{NC}, \emph{CESNA}, and \emph{Louvain}.}}
	\begin{center}
		\begin{tabular}{ |c| c| c| c| c|}
			\hline
			Network & PFCD & NC & CESNA & Louvain \\ \hline
			WorldTrade & 0.52 & 3.097 & 0.38 & 0.04\\ \hline
			PolBlogs & 0.702 & 41 & 2.16 & 0.01 \\ \hline
			Lawyer & 0.765 & 21 & 0.12 & 0.01 \\ \hline
			Malaria & 0.81 & 34 & 1.918 & 0.02 \\ \hline
			WeddellSea & 1.41 & 1461 & 3.18 & 0.03 \\ \hline
			CalTech & 1.497 & 1518 & 4.71 & 0.02 \\ \hline
			DBLP & 2 & 1114 & 0.7 & 0.005 \\ \hline
			Arxiv & 2.31 & 91 & 0.26 & 0.01 \\ \hline
			Rice & 31 & 3476 & 5.4 & 0.33 \\ \hline
			Patent & 446 & -- & 37 & 3.5\\ \hline
			Internet & 2818 & -- & 5400 & 0.63\\ \hline
		\end{tabular}
	\end{center}
\end{table}

{The execution running time of \emph{PFCD} is compared with the state-of-the-art competitors in Table \ref{tbl:execution}, \emph{CESNA} with $O(E)$ and developed in C++, \emph{NC} with $O(|V|^2\times k^2)$ and developed in C, and the fast non-overlapping community detection method, \emph{Louvain}, developed in C and taking $O(|V|log(|V|))$. 
	All of the experiments are performed using a single Lenovo machine with 4GB Ram and Core i5 CPU and the proposed method is implemented in JAVA using JGraphT library. 
	It is worth mentioning that most of the feature-oriented methods are unable to report results in decent time because some of them were developed in MATLAB or require adjacency matrix ($|V| \times |V|$) which is not working on  big networks.
	In the execution time of each method, Louvain algorithm is quite faster compared with the proposed algorithm because first of all its time complexity is  $O(|V|log(|V|))$ which is generally lower than $O(|E|)$  and it is just based on the network structure and does not \textcolor{black}{take into account} node features.
	However, the Louvain algorithm is not able to detect communities on several networks such as CalTech (0.34 F1-score), WorldTrade (0.24 F-score), Arxiv or DBLP (around 0.1 F-score). }


\section{Case study}
\label{sec6}
\begin{figure}[h!]
	\centering
	\begin{minipage}{0.25\textwidth}
		\includegraphics[width=0.7\linewidth]{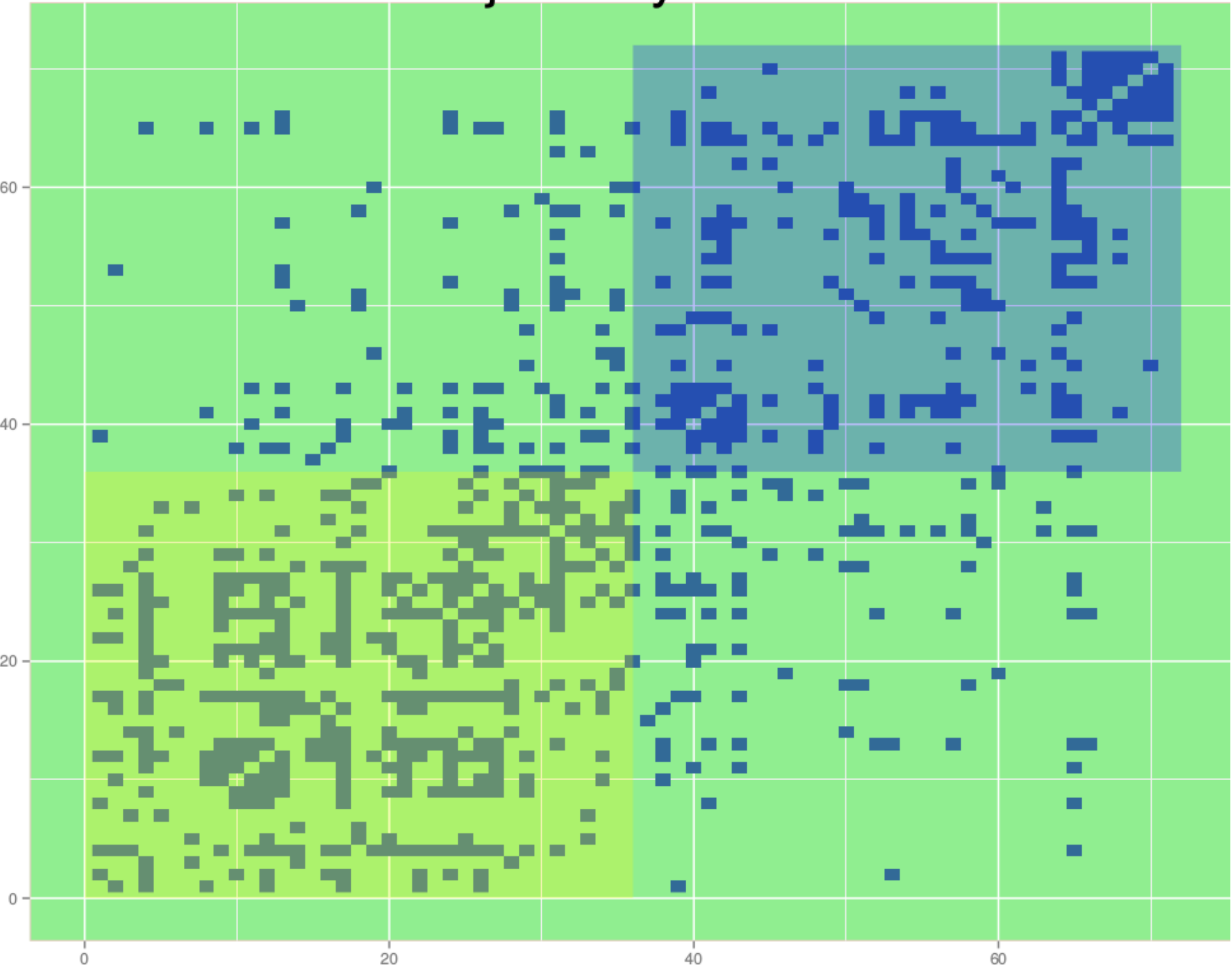}
		\label{fig:LawyerStatus}
		\subcaption{}
	\end{minipage}
	\begin{minipage}{0.25\textwidth}
		\includegraphics[width = 0.7\linewidth]{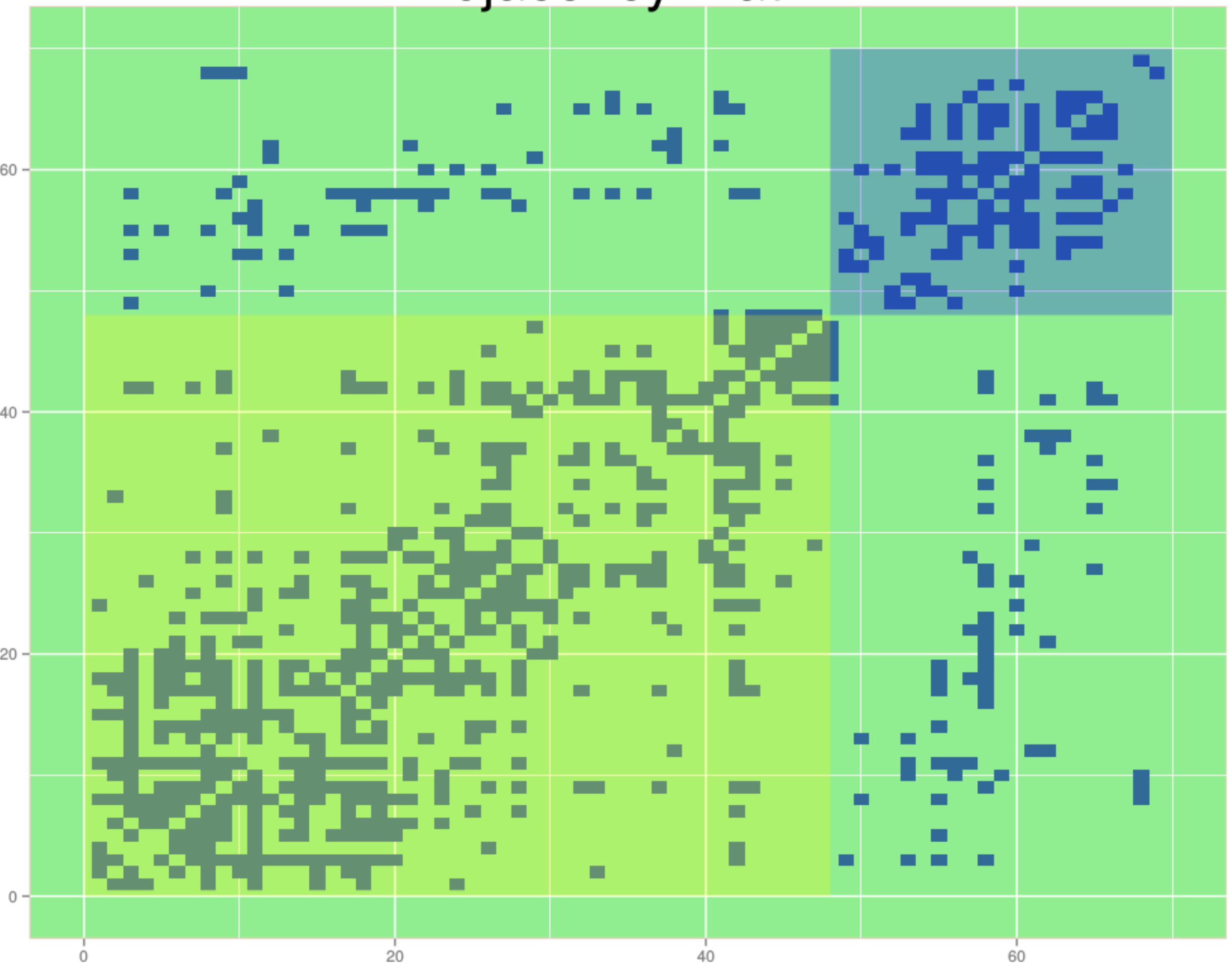}
		\label{fig:LawyerPractice}
		\subcaption{}
	\end{minipage}
	\caption{\label{fig:LawyerCaseStudy}Adjacency matrix according to the value of each feature. (a): status, (b): office. Points represent edges among nodes and each either yellow or blue block shows the group of nodes with a similar feature. }
\end{figure}

Here, we demonstrate the effectiveness of features for detection of communities. \emph{Lawyer} network is considered for further analyzing its communities and the role of features in each community structure. To illustrate the situation, Figure \ref{fig:LawyerCaseStudy} depicts the adjacency matrix sorted by different features, \emph{status (Partner, Associate)} and \emph{office location(Boston, Hartford)}. In Figure \ref{fig:LawyerCaseStudy} part (a), yellow block denotes a group of lawyers with \emph{Partner} status and blue block consists of lawyers with \emph{Associate} status.  {In Figure \ref{fig:LawyerCaseStudy} part (b), lawyers working in \emph{Boston} are shown in yellow block, and the others working in \emph{Hartford} are depicted in blue block.} According to Figure \ref{fig:LawyerCaseStudy}, the features are strong enough to shape communities where nodes with similar features are more likely to share connections together compared with those with different features. The level of strength between the features and communities, is shown in Table  \ref{fig:correlationLawyer}. In the first glimpse, Table \ref{fig:correlationLawyer} shows that the proposed method is able to extract the unique set of features among all possible features for each community based on their strength levels. For example, feature \emph{Partner} has a positive impact on the second community while it does not carry out the same impact on the first one. Moreover, the features, ``lawyers with \emph{associate} status'',  and ``the lawyer's working offices located in \emph{Boston}'', play  important role for shaping the first community. In the same way, ``lawyers with \emph{partner} status''  {whose} working offices located in \emph{Hartford}'' are more influential in shaping the second community. The proposed method is also able to  prioritize the importance levels of specific features of each community. It is shown in Table \ref{fig:correlationLawyer},  \emph{''status''} is more important than \emph{''office location''}.
\begin{table}[t!]
	\caption{the Impact of each feature on each community.}
	\label{fig:correlationLawyer}
	\centering
	\small
	\resizebox{0.5\textwidth}{!}{
		\begin{tabular}{lccccc}
			\toprule[1pt]				
			Community &  Partner & Associate & Boston & Hartford\\ 
			\midrule 
			1 & -0.66 & 0.8 & 0.04 & -0.001\\
			2 & 0.9 & -0.7 &  0 & 0.081\\
			\bottomrule[1pt]    
		\end{tabular} 
	}
\end{table}


\section{Conclusion}
In this work, we introduced a novel graphical model based approach for community detection. The proposed approach, \emph{PFCD}, considered both the network structure and nodal features. The different influence of nodal features on community structures are investigated in our proposed framework. The proposed  model is inferred through an efficient probabilistic  algorithm. The block-coordinate descent algorithm was employed to learn parameters of the model to deal with the latent variables in an efficient computational manner.  In line with the discrimination of features influence on community formation, the priority of each feature on the structure of communities can be inferred from our model. 
The experimental results on synthetic networks justified the strength of the \emph{PFCD} approach on  {detection of} communities compared with the well-known methods.  
Furthermore, a variety of small to large real network datasets  {were} used to evaluate the proposed model based on standard evaluation measures. The results on real networks showed the high performance of the proposed model and very promising results  on the detection of  community structures based on a  network aligned with the nodal features.\par 
There are some future works, such as representation learning approach to derive automatic features from the network structure and extending the proposed method to temporal networks.

\label{sec7}

\ifCLASSOPTIONcompsoc
\else
\fi

\ifCLASSOPTIONcaptionsoff
  \newpage
\fi


%
\vskip 0pt plus -1fil
\begin{IEEEbiography}[{\includegraphics[width=1in,height=1.2in,clip,keepaspectratio]{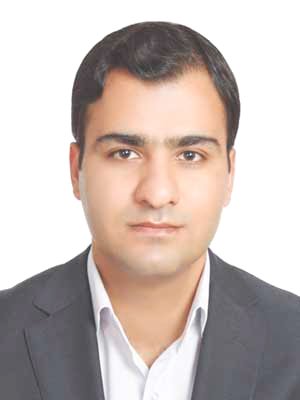}}]{Hadi Zare}
received his masters and the PhD in Statistics from Amirkabir University of Technology in 2007 and 2012, respectively.
He is currently an assistant professor with the faculty of New Sciences and Technologies in the University of Tehran, Tehran, Iran. His main research interests are in Statistical Machine Learning, Probabilistic Graphical Models and Social Network Analysis. More specifically, he has published  papers on structure learning on graphical models, feature selection, exploring  probabilistic models on the summarization of big data structures such as social networks.
\end{IEEEbiography}
\vskip 0pt plus -1fil
\begin{IEEEbiography}[{\includegraphics[width=1in,height=1.2in,clip,keepaspectratio]{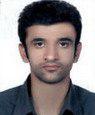}}]{Mahdi Hajiabadi}
	received his masters in Knowledge Engineering and Decision Sciences from University of Tehran in 2016. He's currently working toward on the PhD in Computer Science at the University of Victoria, Victoria, Canada. 
	His main research interests are in Data Mining, Statistical Learning, and Social Networks. He has published some national and international papers on community detection problem in social networks.
\end{IEEEbiography}
\vskip 0pt plus -1fil
\begin{IEEEbiography}[{\includegraphics[width=1in,height=1.2in,clip,keepaspectratio]{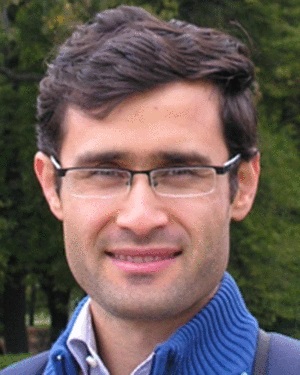}}]{Mahdi Jalili (M’09–SM’16) }
received the B.S. degree in electrical engineering from Tehran Polytechnic, Tehran, Iran, in 2001, the M.S. degree in electrical engineering from the University of Tehran, Tehran, in 2004, and the Ph.D. degree in synchronization in dynamical networks from the Swiss Federal Institute of Technology Lausanne, Lausanne, Switzerland.
He then joined Sharif University of Technology as an Assistant Professor. He is currently a Senior Lecturer with the School of Engineering, RMIT University, Melbourne, Vic., Australia. His research interests include network science, dynamical systems, social networks analysis and mining, and data analytics.
Dr. Jalili was a recipient of Australian Research Council DECRA Fellowship and RMIT Vice-Chancellor Research Fellowship. He is an Associate Editor for the IEEE Canadian Journal of Electrical and Computer Engineering and an Editorial Board Member of the Mathematical Problems in Engineering and Complex Adaptive Systems Modeling.
\end{IEEEbiography}

\end{document}